\documentclass[6pt]{article}

\usepackage[english]{babel}

\usepackage[letterpaper,top=2cm,bottom=2cm,left=3cm,right=3cm,marginparwidth=1.75cm]{geometry}

\usepackage[round, sort]{natbib}
\usepackage{booktabs}

\usepackage{float}

\usepackage{caption}

\usepackage{amsmath}
\usepackage{graphicx}
\usepackage[colorlinks=true, allcolors=blue]{hyperref}
\usepackage{orcidlink} 

\usepackage{multirow}

\usepackage{multicol}

\usepackage{xspace}
\newcommand{\co}{\textit{Control}\xspace}
\newcommand{\ro}{\textit{Recommendation Only}\xspace}
\newcommand{\eo}{\textit{Evidence Only}\xspace}
\newcommand{\rae}{\textit{Recommendation and Evidence}\xspace}
\newcommand{\eva}{\textit{Evaluative AI}\xspace}

\newif\ifanonymous
\anonymousfalse 

\title{An Empirical Examination of the Evaluative AI Framework}
\ifanonymous
    \author{Anonymous Author(s)}
\else
    \author{Jaroslaw Kornowicz \orcidlink{0000-0002-5654-9911}\\
    Paderborn University, Paderborn, Germany\\
    \href{mailto:jaroslaw.kornowicz@uni-paderborn.de}{jaroslaw.kornowicz@uni-paderborn.de}
    }
\fi

\begin{document}
\small
\maketitle

\begin{abstract}
This study empirically examines the ``Evaluative AI'' framework, which aims to enhance the decision-making process for AI users by transitioning from a recommendation-based approach to a hypothesis-driven one. Rather than offering direct recommendations, this framework presents users pro and con evidence for hypotheses to support more informed decisions. However, findings from the current behavioral experiment reveal no significant improvement in decision-making performance and limited user engagement with the evidence provided, resulting in cognitive processes similar to those observed in traditional AI systems. Despite these results, the framework still holds promise for further exploration in future research.
\end{abstract}


\begin{multicols}{2}

\section{Introduction}

In recent years, AI has gained substantial attention for their increasingly sophisticated performance in various applications \citep{Barredo_2020, Rong_2022,albrecht2016gdpr,maccarthy2019examination}. However, their significant limitation compared to simpler methods is their commonly opaque ``black box'' nature, making it difficult to understand how inputs generate outputs \citep{Guidotti_2018}. This is particularly problematic in high-stakes areas like medicine, economics, or law, where understanding the decision-making process is crucial \citep{Rudin_2019}. As a result, the lack of transparency and comprehensibility often leads to distrust and underreliance among potential users, despite the accuracy of these decision-support systems \citep{Jacovi_Marasović_Miller_Goldberg_2021,Mahmud_Islam_Ahmed_Smolander_2022,Zhang_Liao_Bellamy_2020}.

This challenge has spurred the development of several explanatory methods and a surge in interest in Explainable AI (XAI). Initially, it was hoped that XAI would enhance understanding and trust in AI models, thereby improving decision-making quality among users. However, as summarized by recent studies \citep{Lai_Zhang_Chen_Liao_Tan_2023, Schemmer_2022, Vasconcelos_2023,Bertrand_Viard_Belloum_Eagan_Maxwell_2023,Rogha_2023, Schemmer_Kuehl_Benz_Bartos_Satzger_2023}, the results are mixed. While XAI might indeed improve understanding \citep{Ribeiro_Singh_Guestrin_2018}, higher transparency can make models less comprehensible \citep{Poursabzi2021}.  Explanations can improve subjective perception \citep{Bertrand_Viard_Belloum_Eagan_Maxwell_2023}, but also might increase cognitive load \citep{You_Yang_Li_2022, Herm_2023, Ghai2020} and reduce efficiency \citep{Lai_Zhang_Chen_Liao_Tan_2023}. This has led to a situation where users often engage superficially with explanations and develop an overreliance on AI \citep{Chromik_Eiband_Buchner_Krüger_Butz_2021,Bucinca_Malaya_Gajos_2021,Bansal_Wu2021,Chen_Liao_Vaughan_Bansal_2023}, shifting from the original problem of underreliance.

Given that AI is not infallible and often makes better decisions than humans \citep{Mnih_2015, nori2023capabilities}, a calibrated level of trust is essential for a trade-off that encourages user to rely more on AI, while avoiding blind trust \citep{Wischnewski_Krämer_Müller_2023,Vered_Livni2023}. To address the issue of overreliance, various strategies have been developed, such as cognitive forcing functions \citep{Bucinca_Malaya_Gajos_2021} and user-adapted, selective explanations \citep{Lai_Zhang_Chen_Liao_Tan_2023}. This paper discusses another approach to improve human-AI interaction: the ``Evaluative AI'' framework proposed by \cite{Miller_2023}. Critiquing the limited success of existing XAI methods, Miller argues that these methods do not align well with the cognitive processes involved in decision-making. He suggests a paradigm shift from recommender-driven systems to a hypothesis-driven approach, based on the Data/Frame Theory \citep{klein2007data} and abductive reasoning \citep{peirce2009writings}, to better support decision-makers in exploring hypotheses rather than receiving direct recommendation by AI.

This study empirically investigates the effectiveness of the proposed framework in enhancing decision-making by examining its impact on performance, efficiency, and subjective perception. The focus is on one specific element of the framework: offering evidence \textit{for and against} potential option without providing direct recommendations. Rather than giving a recommendation and explaining it, the framework refrains from making any recommendations. Instead, it offers evidence supporting and opposing each option, which is only displayed if requested by the decision-maker. This studies research question is:

\begin{itemize}
    \item \textbf{RQ:} Can a decision support system that offers evidence for and against potential options, without providing direct recommendations, improve the decision-making process?
\end{itemize}

Currently, only three studies directly apply Miller's framework: \cite{Castelnovo_Crupi_Mombelli_Nanino_Regoli_2023} developed a contrastive explanation technique for ranking classifications, and \cite{Le_Miller_Zhang_Sonenberg_Singh_2024} created a tool for image classification, though neither has undergone empirical testing. 

During the development of the present study, an empirical evaluation by \cite{Le_Miller_Sonenberg_Singh_2024} was conducted, comparing a hypothesis-driven approach with recommendation-driven and explanation-only methods. They found that the hypothesis-driven approach improved decision quality without increasing decision time, and participants cognitively engaged with the evidence, thereby considering the uncertainty of the underlying models. This current study differs in several respects. Compared to \cite{Le_Miller_Sonenberg_Singh_2024}, the task here is significantly more objective and realistic for participants. While their task involved classifying a subjective house price into low, medium, or high using six features, the task in this study is to estimate whether an income is above or below the median based on 20 features.

This study provides a more detailed picture, as it includes a control group without any AI assistance and a group that receives both recommendations and evidence. Another difference lies in the incentive design; in this study, more incentive per task was offered to simulate a higher-stakes situation. In a pretest, it was found that evidence presented in bar chart format (as used in \cite{Le_Miller_Sonenberg_Singh_2024}) was not well understood, so textual descriptions of the evidence were added here. Lastly, in \cite{Le_Miller_Sonenberg_Singh_2024} experiment, low-level evidence was shown by default, which could potentially lead to anchoring effects and influence the decision-making process. In this study, no evidence is shown by default, allowing decision-makers the freedom to choose and gives further opportunities for behavioral analysis.

The results of the present study paint a different picture than those of \cite{Le_Miller_Sonenberg_Singh_2024}. Overall, the findings indicate that the ``Evaluative AI'' framework in this experiment did not improve decision-making performance. They also reveal that participants engaged only superficially with the provided pro and con evidence, despite all AI systems influencing the decision-making processes leading potentially to cognitive offloading.

\section{Background and Related Work}

The concept of developing explainability methods based on decision-making processes to create more human-centered XAI is not entirely novel. According to \cite{Vered_Livni2023} XAI researchers fail to align explanations with the human reasoning process. \cite{Vasconcelos_2023} analyze the problem of overreliance from a cost-benefit tradeoff perspective. According to their framework, overreliance can result from a strategic decision in which users weigh the value of engaging with a recommendation and its explanation against the potential benefits. \cite{Miller_2019} advocated for an interdisciplinary approach by aligning with established knowledge about explanations in disciplines like philosophy and psychology. He posited that explanations in XAI should be primarily contrastive, selective, and tailored to fit the social context. 

\cite{Wang_Yang_Abdul_Lim_2019} also developed a XAI framework, drawing on prior research in human decision-making. A key aspect of their framework is its emphasis on forward reasoning, as informed by the hypothetico-deductive model, contrasting with backward reasoning approaches \citep{popper2014conjectures,croskerry2009universal}. This methodology suggests that forming hypotheses based on available information (forward reasoning) is more effective than initially devising hypotheses and then seeking confirmation within the data (backward reasoning). In this context, recommender-based XAI systems align more closely with backward-oriented reasoning, as they present recommendations directly to the user as initial hypotheses. \cite{Gouveia_2024} contend that most AI systems currently lack the ability to provide explanations based on abductive reasoning. However, they suggest that Large Language Models (LLMs) could become valuable in this regard in the future.

\cite{Miller_2023} aims to initiate a paradigm shift towards hypothesis-driven XAI with his framework. He believes that recommender-driven XAI is not aligned with cognitive thinking processes and thus limits agency, which can be crucial in medium/high-stakes and low-frequency decisions. 
He is motivated by, in his opinion, the disappointing results of previous XAI systems on the decision-making process. This may be because users engage minimally with the explanations. Some researchers have addressed this issue and proposed several solutions. Notably, the cognitive forcing functions by \cite{Gajos_Mamykina_2022} and \cite{Bucinca_Malaya_Gajos_2021}, which, while increasing cognitive engagement, do not meet \cite{Miller_2023} criteria for a good decision support system.

The framework is built on several decision research theories. For instance, \cite{Miller_2023} uses ``cardinal decision issues'' \citep{yates2012evidence} to define what a good decision support system should look like. It should help identify options and narrow the decision space, identify possible outcomes, assess the probabilities and consequences of these outcomes, and assist in finding a trade-off, making this understandable for the user. These criteria are connected with theories about cognitive processes during decision-making, especially focusing on abductive reasoning --- the process of forming hypotheses and assessing their probabilities to explain observations. Furthermore, \cite{Miller_2023} connects this with \cite{klein2007data} findings, that decision-makers initially intuitively narrow the decision space and then go through all remaining options, seeking pros and cons for the options. 

XAI systems built on this framework should not provide recommendations (e.g., the patient has disease A). Instead, they should highlight the most likely options (referred to as hypotheses), such as indicating that diseases A and C are the most probable. Additionally, they should support the decision-maker in exploring these options by providing for and against evidence (e.g., it could be disease A because..., but against this is...). Most importantly, the decision-maker should have the autonomy to decide which options to investigate and when.

Several studies address these elements. \cite{Cresswell_Sui_Kumar_Vouitsis_2024} utilized conformal prediction to identify the most likely options in an image classification problem, demonstrating that this approach improved decision accuracy. \cite{Lai_Tan_2019} showed that heatmaps as text classification explanations (although they are not, per se, pro and con arguments according to the framework) slightly improve user performance with further gains when combined with recommendations, achieving the best results. Similarly, \cite{Lai_Liu_Tan_2020} highlighted the positive impact of explanations alone but found no additional benefit from integrating recommendations. On the other hand, \cite{Carton_Mei_Resnick_2020}, using an AI that performs worse than humans, found no benefits from explanations, recommendations, or their combination. \cite{Bucinca_Malaya_Gajos_2021} experimented with a similar approach where decision-makers could receive recommendations alongside explanations either on demand or after a waiting period. This method decreased overreliance but did not improve performance compared to a baseline XAI condition. \cite{Gajos_Mamykina_2022} found that providing an explanation without the recommendation led to better decisions and learning gains. \cite{Ma_Lei_Wang_Zheng_Shi_Yin_Ma_2023} showed that presenting recommendations when the AI is more likely to be correct for a specific observation, while still providing the explanation, encouraged participants to think more independently, resulting in lower overreliance. \cite{Spatola_2024} on other hand, found that explanatory guidance by an AI chatbot did not improve outcomes and that users are often focused on efficiency, but risk over-assimilation, that can lead to lower performance in the long term. Finally, as described detaily, \cite{Le_Miller_Sonenberg_Singh_2024} demonstrated that presenting evidence for multiple hypotheses while hiding the recommendation can increase decision accuracy.

\section{Method}

\subsection{Overview}

To answer the research question of whether the framework's element regarding \textit{evidence for} and \textit{evidence against} can improve the decision-making process, an incentivized between-subjects experiment was conducted online. A mixed methods approach is used to quantitatively assess the participants' behavior and qualitatively understand how they decided. Participants were asked to probabilistically estimate, based on the 20 personal characteristics of four individuals, whether each of them earned a net income above the population median. To do this, participants were instructed to estimate the probability as a percentage of how likely it was that this was the case. The participants received a fixed payout of £3, along with a performance-based bonus --- the more accurate their estimates, the higher their bonus payout.

Income estimation is a common task in research studies because it is simple for participants to understand \citep{Ma_Lei_Wang_Zheng_Shi_Yin_Ma_2023,Zhang_Liao_Bellamy_2020}. This task is especially useful when large sample sizes are needed, as recruiting experts in a specific field can be challenging. Therefore, laypeople are often recruited instead.

Since \cite{Miller_2023} sees the Evaluative AI framework as applicable in medium/high-stakes and low-frequency decisions, the number of tasks participants were required to complete was intentionally set to four. While similar studies use more tasks (such as 12 \citep{Le_Miller_Sonenberg_Singh_2024} or 14 \citep{Bucinca__2024_contrastive}), this study aims to create a higher-stakes situation artificially by offering a relatively high potential bonus payment per task.

A simple logistic regression is used as the AI model. The output of the trained model is shown as a recommendation. SHAP (SHapley Additive exPlanations) \citep{shap_paper} was used to generate pro and con evidence, classifying individual personal characteristics into supporting or opposing arguments. For better understanding, SHAP values were displayed both graphically and as text.

To empirically test whether AI, based on the evaluative AI framework as proposed, can improve decision quality, participants were randomized into five groups. In the first group, which served as the control group, participants worked without any assistance, while in the other four groups, participants received different forms of AI as decision support.

The specific treatment groups are explained in the section on \nameref{sec:method_conditions}, followed by a description of the hypotheses metrics used for the empirical evaluation in \nameref{sec:method_hypotheses}. The experiment's procedure, from the participant's perspective, is detailed in \nameref{sec:mathod_procedure}. The sections on \nameref{sec:method_dataset} and \nameref{sec:method_model} describe the dataset used for the task and the AI developed. Finally, the recruitment of participants is discussed in \nameref{sec:method_participants}.
The experiment was preregistered before data collection\footnote{\href{https://osf.io/k2jhf}{https://osf.io/k2jhf}}. The ethics board of the University of Paderborn approved the research project.

\subsection{Experimental Conditions}
\label{sec:method_conditions}
The participants in the experiment were randomly assigned to one of the following groups:
\begin{itemize}
    \item In the \co group, participants completed the task alone without any assistance.
    
    \item In the \ro group, participants received only AI recommendations.\\
    Below the features and the input field for the estimated probability, the AI assistance was displayed: \textit{``The AI suggests a probability of x\%''}.
    
    \item In the \eo group, participants received all evidence for and against each option directly.\\
    Evidence for and against was presented side by side. At the top, a bar chart with the normalized SHAP values and feature values was displayed, and below that, a text describing the AI's evidence was shown.
    
    \item In the \rae group, the AI resembled a classical XAI system where both the recommendation and the evidence were displayed directly.\\
    In this case, the recommendation is displayed at the top, with the pro and con evidence shown below it.
    
    \item The \eva group, that represented the framework, is similar to the \eo group, but participants do not receive the evidence directly, but choose when to view it.
    Two buttons were displayed, allowing participants to view the pro and con evidence separately. The possible click times were tracked.
    
\end{itemize}

Multiple screenshots of the experimental interface can be found in the Apendix \nameref{appendix:screenshots}.

\subsection{Hypotheses and Dependent Variables}
\label{sec:method_hypotheses}
Decisions and decision-making processes can be evaluated in various ways \citep{Lai_Chen_Smith-Renner_Liao_Tan_2023}. This study follow previous work and primarily focus on the performance of the decisions, meaning how good the decisions are, the efficiency, meaning how much time is required to make the decisions and cognitive load, meaning how much cognitive effort is required for the decision. Based on the framework, the first hypothesis is:

\begin{itemize}
\label{h1}
\item \textbf{H1:} The \textbf{best decisions} will be made in the \eva group.
\end{itemize}

The performance of these probabilistic estimations will be evaluated using the Brier score, which considers the estimated probabilities and the actual incomes. This approach was also used by \cite{Le_Miller_Sonenberg_Singh_2024}. Compared to simple binary yes-or-no decisions, probabilistic responses allow for directly measuring participants' confidence, thereby providing a more detailed answer. Let $p_i$ be a participant's estimated probability that the income of individual $i$ is above the median, and let $o_i$ be the actual outcome, where $o_i = 1$ if the income is above the median and $o_i = 0$ if the income is below the median. The Brier score is then defined as:

$$
\frac{1}{N} \sum_{i=1}^{N} (p_i - o_i)^2
$$

where $N$ is the total number of individuals in the task. Thus, the better the assessments, the lower the score.

Based on their Brier score, participants receive a bonus payment as a monetary incentive. With random guessing—always indicating 50\% --- the Brier score would be 0.25. At this score (and above), the bonus payment is £0. The lower the Brier score, the higher the payout, up to a Brier score of 0, which corresponds to £6.

\begin{itemize}
\label{h2}
\item \textbf{H2:} The \textbf{slowest decisions} will be made in the \eva group.
\end{itemize}

It is also plausible to assume that such a AI system will require more time. On one hand, participants in \eva might choose to forgo the assistance entirely or partially, which should lead to a shorter decision time compared to the condition where the pros and cons are fully visible from the start. On the other hand, in \eva, there is an additional decision on whether to view the evidence, which takes time. The dependent variable, time, is calculated here as the average time participants need to complete the tasks.

\begin{itemize}
\label{h3}
\item \textbf{H3:} The \textbf{highest cognitive load} will be observed in the \eva group.
\end{itemize}

As speculated by \cite{Miller_2023}, such a decision aid system requires more effort from the user. Therefore, it can be expected that the cognitive load will increase. This is similar to the hypothesis regarding time; although there is less information available at the start, there are more decisions for the participant to make. Cognitive load is assessed subjectively using the NASA-TLX scale \citep{Hart_2006, Schuff_Corral_Turetken_2011}. This is done once after the tasks are completed.

\subsection{Decision-Making Process}
To understand how the participants arrived at their decisions within the experimental task, a qualitative component was added. After completing the task, participants were asked to describe their decision-making process in words. The exact question was: \textit{``Please describe your decision-making process for the previous estimates. How did you make your decisions?''} This was a mandatory field. For the analysis, qualitative content analysis was used to classify the responses. This allows us to quantify the statements and identify further differences between the treatments \citep{Mayring_2015}. The classification was carried out by the author with the assistance of the LLM GPT-4o \citep{Chew_Bollenbacher_Wenger_Speer_Kim_2023, Tai_Bentley_Xia_Sitt_Fankhauser_Chicas-Mosier_Monteith_2024}.

\subsection{Procedure}
\label{sec:mathod_procedure}
\textbf{Start.} The experiment was conducted online using oTree software \citep{chen2016otree}. Participants were recruited through Prolific.com and directed to the experiment via the platform. They were first required to enter their Prolific ID, read the privacy policy, and then complete a survey on demographic data.

\textbf{Introduction.} The participants were randomly assigned to one of the five treatments. The study began with a general instruction (see \autoref{appendix:instructions}). Participants were then given 5 comprehension questions (4 in control condition), with a maximum of two incorrect responses allowed per question. If participants incorrectly answered at least one of these comprehension questions three times, they were disqualified from continuing the study. In such cases, participants were instructed to return their submissions to the Prolific website, and their data were excluded from subsequent analyses. The instructions and questions were structured according to the treatment. Next, the explanation of the personal characteristics of the individuals to be assessed within the experimental task was provided. In addition to the explanation, the average values of the features were also displayed. The instructions and the explanation of these characteristics could be accessed during the task.

\textbf{Experimental Task.}
Participants were introduced to four individuals one after the other and, based on their personal characteristics, were asked to estimate the likelihood (in percentage) that each individual earned a net income above the median. Participants adjusted their percentage estimate using a slider, which was initially set to a default value of 50\%. Participants received feedback on their estimates only after the fourth round.

Depending on the treatment, the AI assistance was displayed below if available, the recommendation was shown first, followed by the pros and cons on the left and right sides, respectively.

\textbf{Desicion-Making Process}. After completing the tasks, participants were asked to describe their decision-making process during the task. For this purpose, a mandatory free-text field without a character minimum or maximum was provided.

\textbf{NASA-TLX}. Finally, participants completed the NASA-TLX questionnaire.

\subsection{Dataset}
\label{sec:method_dataset}
While many previous studies that also used income estimates relied on the widely used \textit{adult dataset} \citep{adult_2}, a new dataset was compiled for this study. This dataset is more recent and includes additional variables.

The data used comes from the SOEP dataset \citep{goebel2019german}. The sample for this experiment includes 7,708 individuals, all of whom are neither retired nor unemployed. The SOEP data can be requested from the German Institute for Economic Research (DIW), and the code for generating the dataset and model is available in the \href{https://osf.io/7pbt2/?view_only=12447fb8850a4ba68628df16b4536a9b}{public repository}.

The dataset was divided into a training sample, which was used to train the AI (logistic regression), and a test sample, which was used to evaluate the AI. From the test sample, an experimental sample of 20 individuals was randomly selected (under the condition that the AI performs similarly on this sample as it does on the entire test sample and that the sample is sufficiently diverse). From these 20 individuals, 4 were randomly assigned to participants for the experimental task.

The dataset contains the following variables: \textit{Body weight}, \textit{Body height}, \textit{Is male}, \textit{Age}, \textit{Has part-time work}, \textit{Work change last year}, \textit{Time pressure at work}, \textit{Sick days last year}, \textit{Number of children}, \textit{Married}, \textit{Divorced}, \textit{Smoking}, \textit{Drinks alcohol}, \textit{Eats meat}, \textit{Student or PhD}, \textit{Has university degree}, \textit{Health status}, \textit{Interested in politics}, \textit{Health satisfaction}, \textit{Life satisfaction}.

\subsection{Regression Model for Income Assessment}
\label{sec:method_model}
For this task, a simple logistic regression model was used as AI (more complex learning algorithms, such as XGBoost, did not lead to any significant improvement). The trained model achieved an ROC AUC of 0.85 and a Brier score of 0.155 on a test dataset. In the experimental sample, a ROC AUC of 0.83 and a Brier score of 0.178 was obtained. If one were to use a decision threshold of 50\%, one would be correct in 15 of 20 cases. 

The model's generated class probabilities for each individual were used as recommendations. For and against evidence is based on SHAP. SHAP can generate feature-based and local explanations for the output of models. In this case, for each individual being assessed, it generates a value for each characteristic, indicating the extent to which that characteristic contributes to the output. Positive contributions are considered positive evidence, while negative contributions are considered negative evidence. These contributions are displayed separately in bar charts.

\begin{figure*}[]
    \centering
    \includegraphics[width=0.7\linewidth]{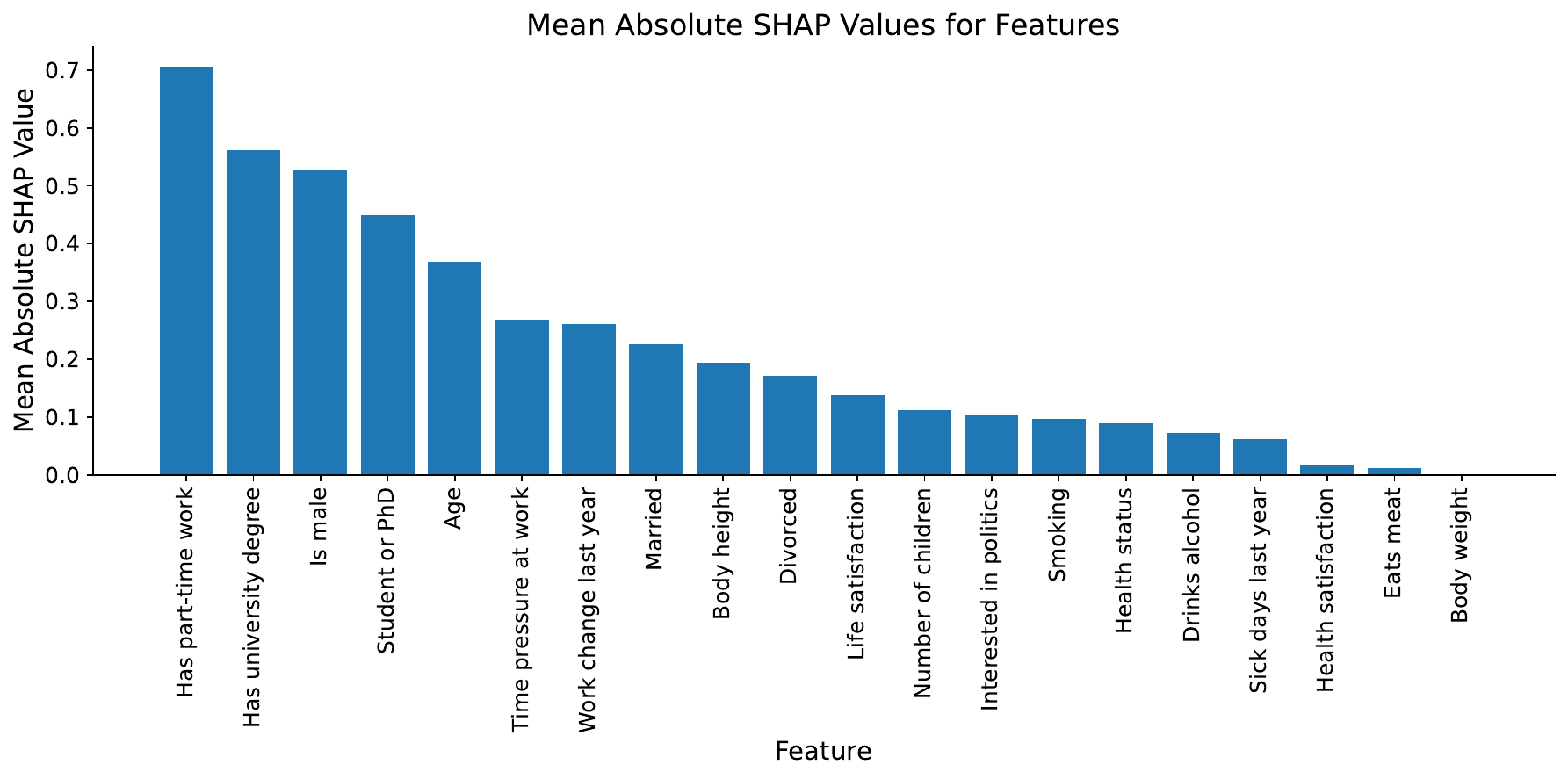}
    \captionsetup{width=0.9\linewidth}
    \caption{Mean Absolute SHAP Values for Features in Model Interpretation. The horizontal bar chart ranks features based on their mean absolute SHAP values, which indicate the average impact of each feature on the model’s predictions.}
    \label{fig:mean_shap}
\end{figure*}

\autoref{fig:mean_shap} shows the average absolute SHAP values of the features across the 20 individuals in the experimental sample.

Similar to \cite{Bucinca__2024_contrastive}, the SHAP-based pro and con evidence were converted into text form and displayed below the bar charts. The SHAP values and the actual values were taken into account in this process. Since the dataset was standardized, the features were comparable. The LLM GPT-4o was used to convert the numerical pro and con evidence into text (the code can also be found in the online appendix).

\subsection{Participants}
\label{sec:method_participants}
The studies were conducted in October 2024. Participants were recruited from the platform Prolific.com. The Paderborn University Institutional Review Board approved the study.

Before recruiting participants, the required sample size was computed in a power analysis for a ANOVA using G*Power \citep{gpower}. To correct for testing multiple hypotheses, a Bonferroni correction was applied. The default effect size $f = 0.25$ (i.e., indicating a medium effect) was specified, with a significance threshold $\alpha = 0.005$ (i.e., due to testing multiple hypotheses), a statistical power of $(1 - \beta) = 0.9$, and the investigation of 5 different experimental conditions/groups. This resulted in a required sample size of 375 participants for the study.

Since the SOEP data used in this study comes from the German population, only participants from Germany were recruited. Additionally, the study was conducted in German, which meant that only participants who are fluent in German were recruited. To ensure high-quality participation, only participants with an approval rating of over 95\% and who had completed at least 50 studies were selected.

\section{Results}
The collected experimental data (excluding participants' personal data) and the analysis codes are available in the online appendix. The analysis was conducted using Python with various packages, and the complete list with version numbers is also available in the online appendix. All p-values reported here were adjusted using the Bonferroni correction.

For the experiment, a total of 439 participants were initially recruited. Of these, 21 were excluded due to failing the comprehension questions, and 42 others voluntarily withdrew at various points. One participant was removed because they did not provide an answer to the question about their decision-making process. This resulted in the final number of 375 participants, matching the number required according to the power analysis.

250 (66\%) of the participants were male, and the average age was 32.7 years. On average, participants received a bonus payment of £1.79. The distribution of participants across the groups was not entirely even: there were 62 participants in \co, 77 in \ro, 64 in the \eo, 81 in \rae, and 91 in \eva.

\subsection{Decision Performance} 
\begin{figure*}[]
    \centering
    \includegraphics[width=0.6\linewidth]{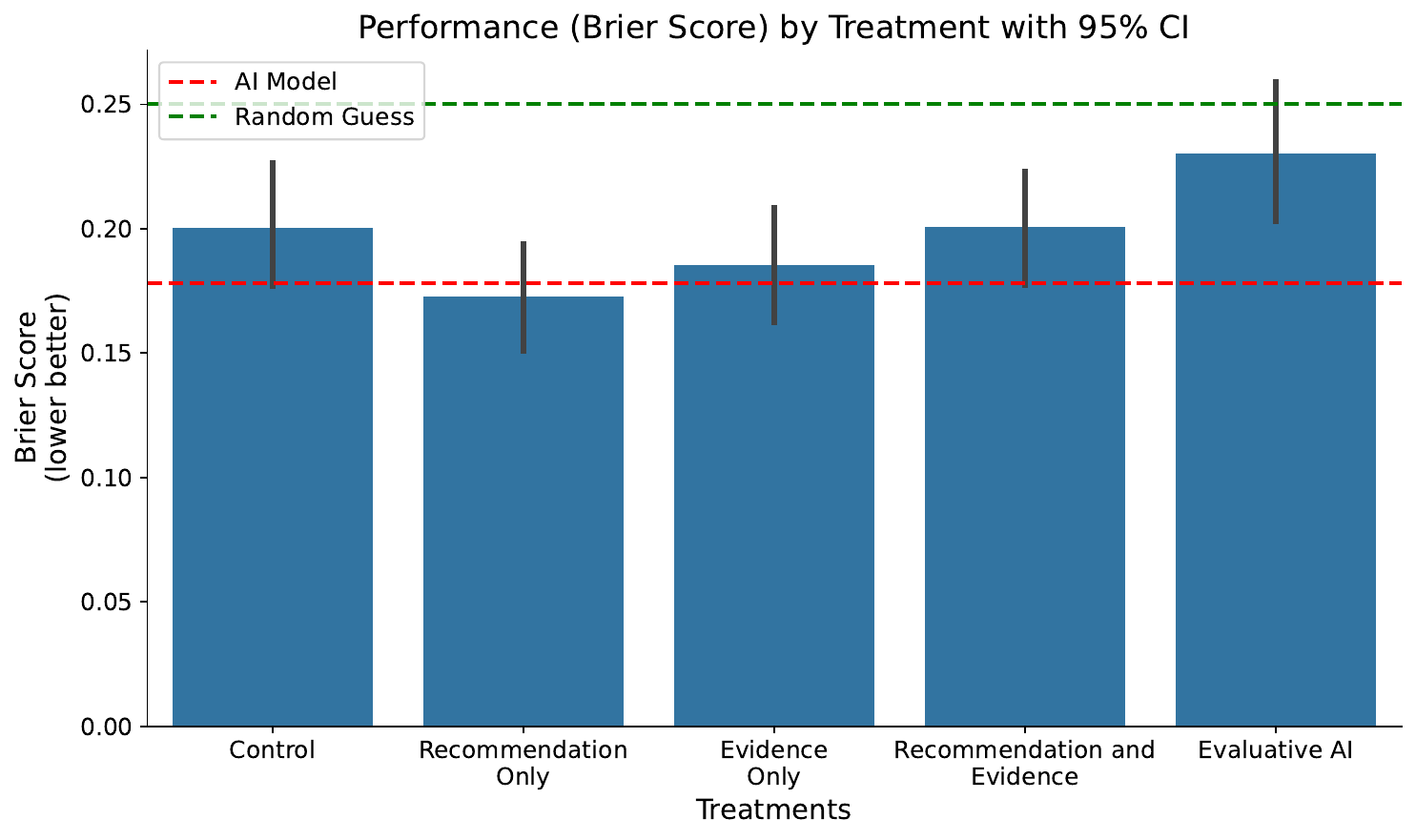}
    \captionsetup{width=0.9\linewidth}
    \caption{Comparison of Brier Scores Across Different Treatments. The bar chart presents the Brier score performance (lower values indicate better predictive accuracy) for various treatment groups: \co, \ro, \eo, \rae, and \eva. Error bars denote the 95\% confidence intervals. Horizontal dashed lines indicate benchmarks for AI Model (red) and Random Guess (green) performance. The results suggest similar performance levels across treatments, with no significant deviations observed.}
    \label{fig:brier_score}
\end{figure*}

Brier score is used to determine the decision performance --- the better the estimates, the lower the score. \autoref{fig:brier_score} illustrates the average Brier scores per treatment with 95\% confidence intervals. While random guessing would result in a score of 0.25 and the logistic regression on the experimental sample achieved a score of 0.178, only the participants in \ro performed better on average ($M = 0.173$, $SD = 0.098$). The second best was \eo ($M = 0.185$, $SD = 0.09$), followed by \co ($M = 0.2$, $SD = 0.098$) and \rae ($M = 0.201$, $SD = 0.115$), with \eva being the lowest ($M = 0.23$, $SD = 0.139$). The statistical testing of the differences for the first hypothesis followed the analysis steps proposed by \cite{Sawyer_2009}. The Shapiro-Wilk test indicated that the data were not normally distributed, so the non-parametric Kruskal-Wallis test was used. According to this test, there is no significant difference between the groups in terms of the Brier score ($p=0.154$), and therefore, H1 is rejected.

\subsection{Decision Time} 
\begin{figure*}[]
    \centering
    \includegraphics[width=0.6\linewidth]{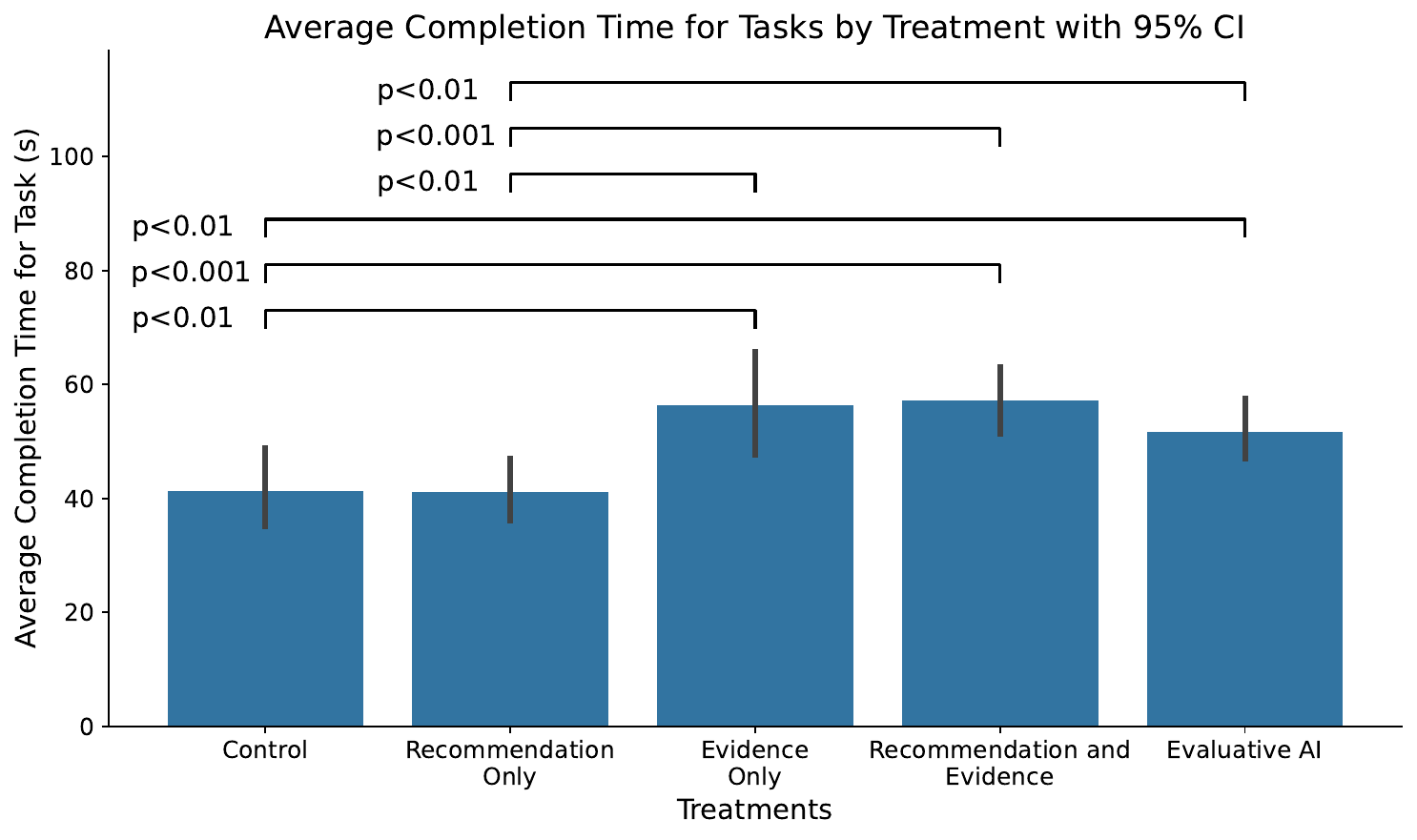}
    \captionsetup{width=0.9\linewidth}
    \caption{Average Task Completion Time Across Different Treatments. This bar chart shows the average time taken (in seconds) to complete tasks for each treatment group: \co, \ro, \eo, \rae, and \eva. Error bars represent the 95\% confidence intervals. Significant differences between groups are indicated by p-values above the bars, illustrating where statistically significant differences ($p<0.01$, $p<0.001$) were observed between treatments.}
    \label{fig:time}
\end{figure*}

Decision time was measured as the average time participants took from the start of a task to the submission of their estimate. There were no major outliers that needed to be removed from the data. \autoref{fig:time} illustrates the average decision times in seconds per treatment with 95\% confidence intervals. Significance bars indicate significant differences between the treatments. Participants in \ro  ($M = 41.198$, $SD = 24.58$) and in \co ($M = 41.343$, $SD = 27.458$) were the fastest, followed by \eva with a larger difference ($M = 51.736$, $SD = 25.915$), \eo ($M = 56.406$, $SD = 26.997$), and \rae ($M = 57.185$, $SD = 27.599$). The tests for significance followed the same steps as for the first hypothesis. Again, the Shapiro-Wilk test indicated that the data were not normally distributed. The Kruskal-Wallis test showed that significant differences exist between the groups ($p < 0.001$). Dunn's post hoc test indicated significant differences between \co and \eo ($p < 0.01$), \rae ($p < 0.001$), and \eva ($p < 0.01$), as well as between \ro and \eo ($p < 0.01$), \rae ($p < 0.001$), and \eva ($p < 0.01$). Although there are significant differences between the treatments, H2 is also rejected.

\subsection{Cognitive Load} 
\begin{figure*}[]
    \centering
    \includegraphics[width=0.6\linewidth]{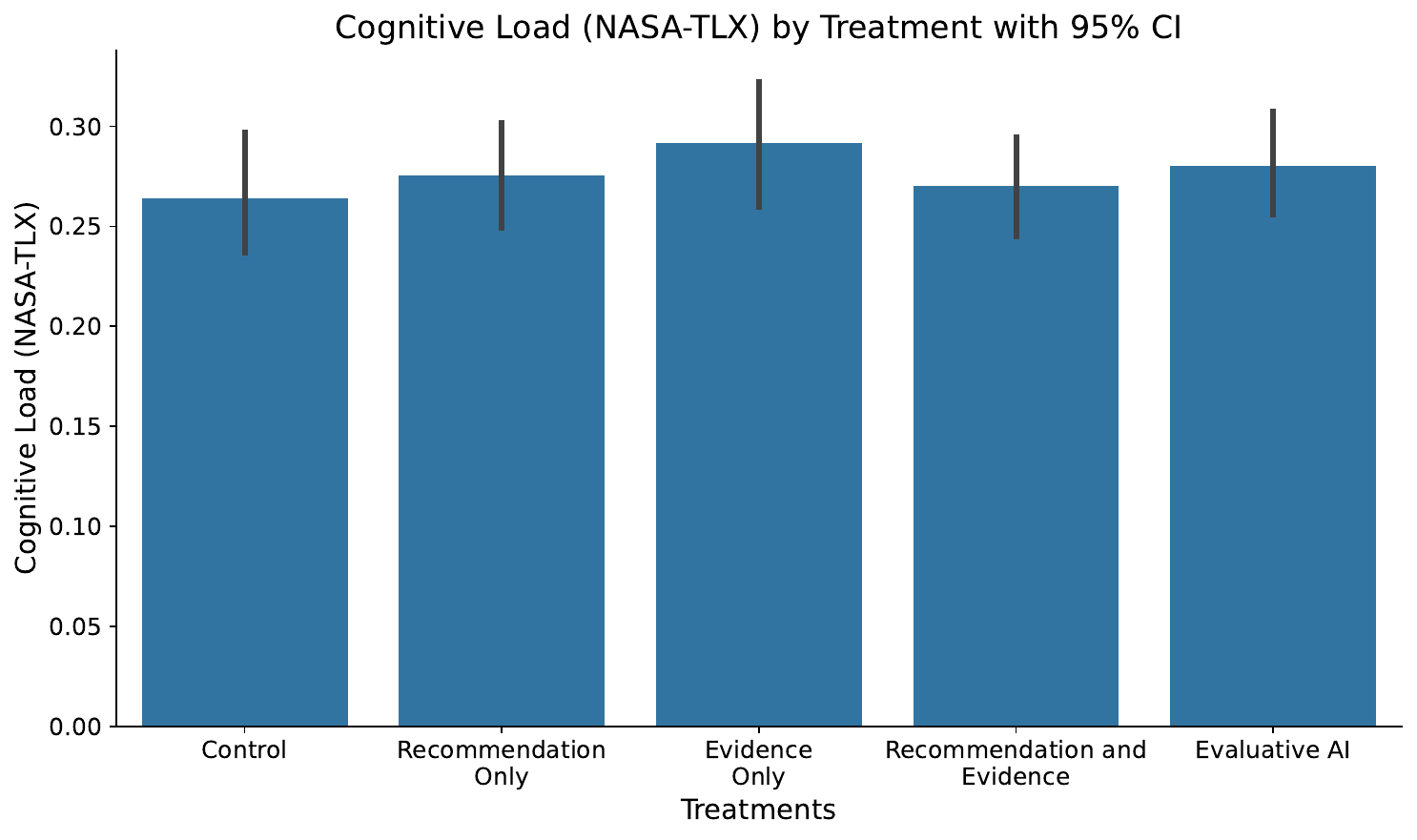}
    \captionsetup{width=0.9\linewidth}
    \caption{Cognitive Load (NASA-TLX) Across Different Treatments. The bar chart illustrates the cognitive load scores, as measured by the NASA Task Load Index (TLX), for each treatment group: \co, \ro, \eo, \rae, and \eva. Error bars show the 95\% confidence intervals, providing an indication of the variability within each group. The results suggest similar cognitive load levels across treatments, with no significant deviations observed.}
    \label{fig:nasa_score}
\end{figure*}

Cognitive load was assessed subjectively using the NASA-TLX scale, and the average values with confidence intervals are shown in \autoref{fig:nasa_score}. Participants experienced the lowest average cognitive load in \co ($M = 0.264$, $SD = 0.12$), followed by \rae ($M = 0.27$, $SD = 0.12$), \ro ($M = 0.275$, $SD = 0.119$), \eva ($M = 0.28$, $SD = 0.125$), and \eo ($M = 0.292$, $SD = 0.126$). The Shapiro-Wilk test indicated that the data were not normally distributed, and the Kruskal-Wallis test showed no significant differences between the treatments. Therefore, H3 is also rejected.

\subsection{Decision-Making Process} 
\begin{figure*}[]
    \centering
    \includegraphics[width=0.6\linewidth]{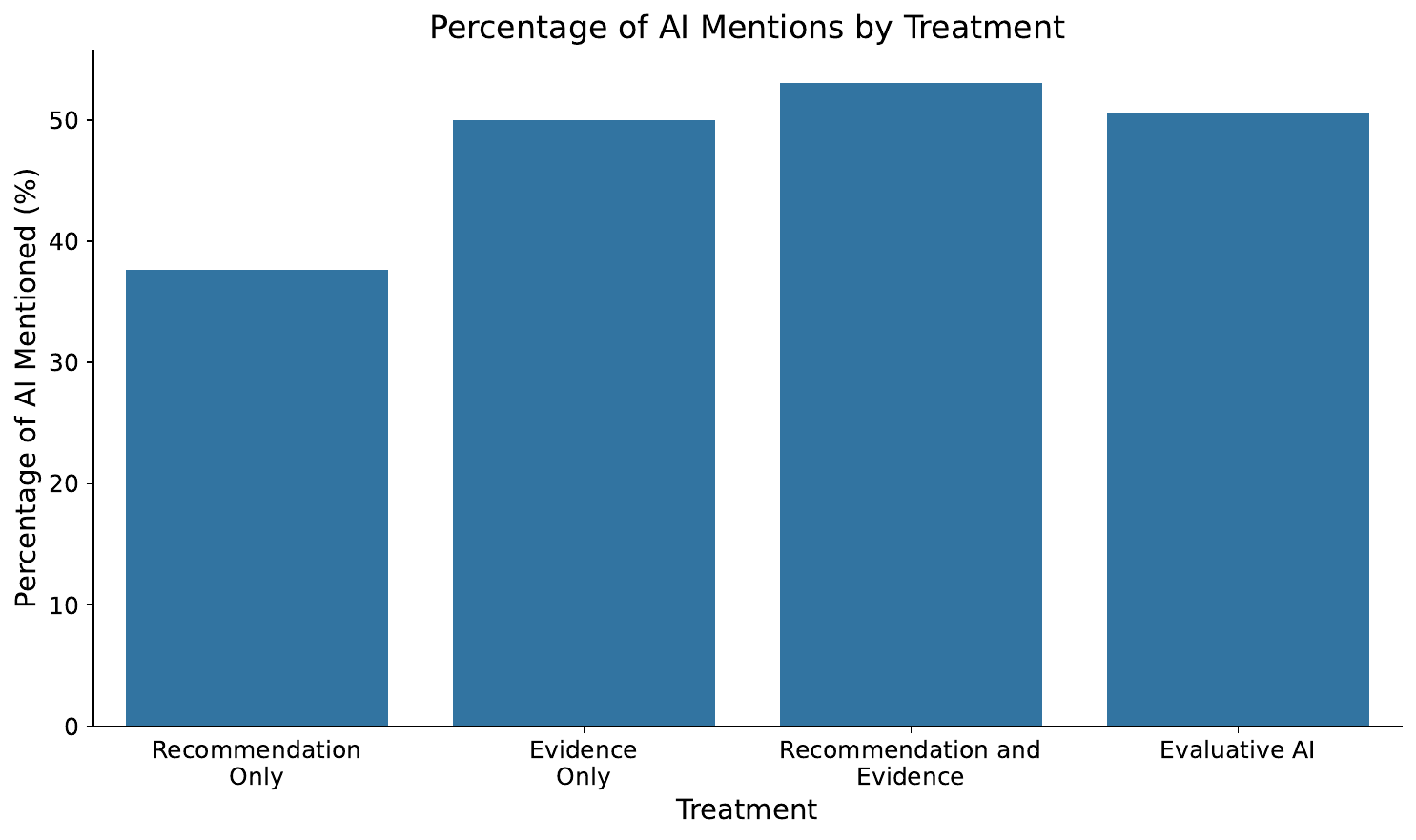}
    \captionsetup{width=0.9\linewidth}
    \caption{Frequency of AI Mentions by Treatment Group. This bar chart displays the percentage of participants mentioning AI across four treatment groups: \ro, \eo, \rae and \eva.}
    \label{fig:ai_mentioned_frequency}
\end{figure*}

After completing all four tasks, the experiment participants were asked how they arrived at their decisions. \autoref{fig:ai_mentioned_frequency} shows the percentage of times participants in each treatment group (excluding \co) mentioned the AI in their decision-making process. Although AI was mentioned the least in \ro ($37.66\%$) compared to \eo ($50\%$), \rae ($53.09\%$), and \eva ($50.55\%$), this difference is not statistically significant according to a pairwise chi-squared tests.

\begin{figure*}[]
    \centering
    \includegraphics[width=0.6\linewidth]{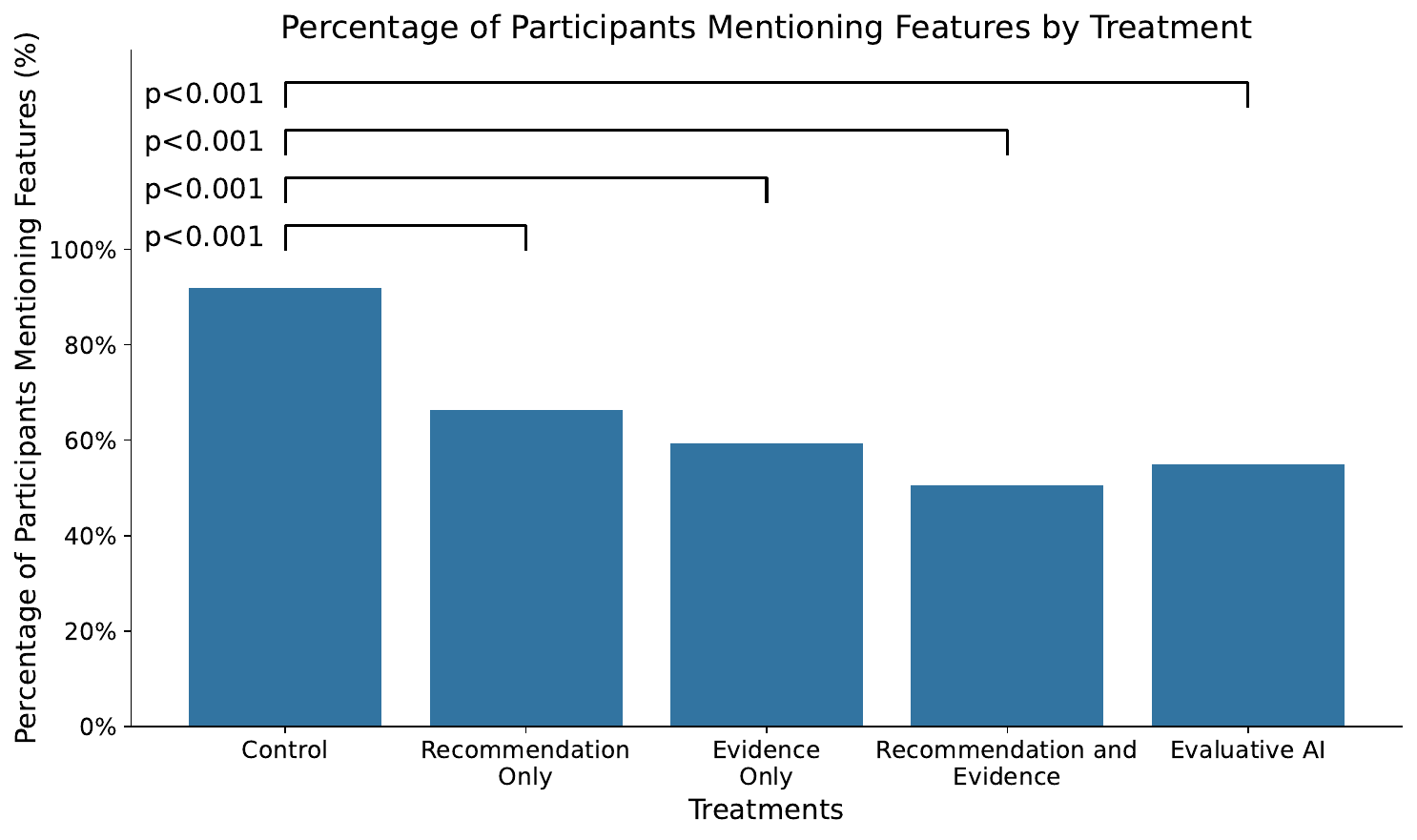}
    \captionsetup{width=0.9\linewidth}
    \caption{Proportion of Participants Mentioning Features by Treatment Group. The bar chart illustrates the proportion of participants mentioning specific features across five treatment groups: \co, \ro, \eo, \rae, and \eva. Significant differences between the Control group and other treatments are marked with p-values ($p<0.001$).}
    \label{fig:features_proportion_mentioned}
\end{figure*}

\begin{figure*}[]
    \centering
    \includegraphics[width=0.6\linewidth]{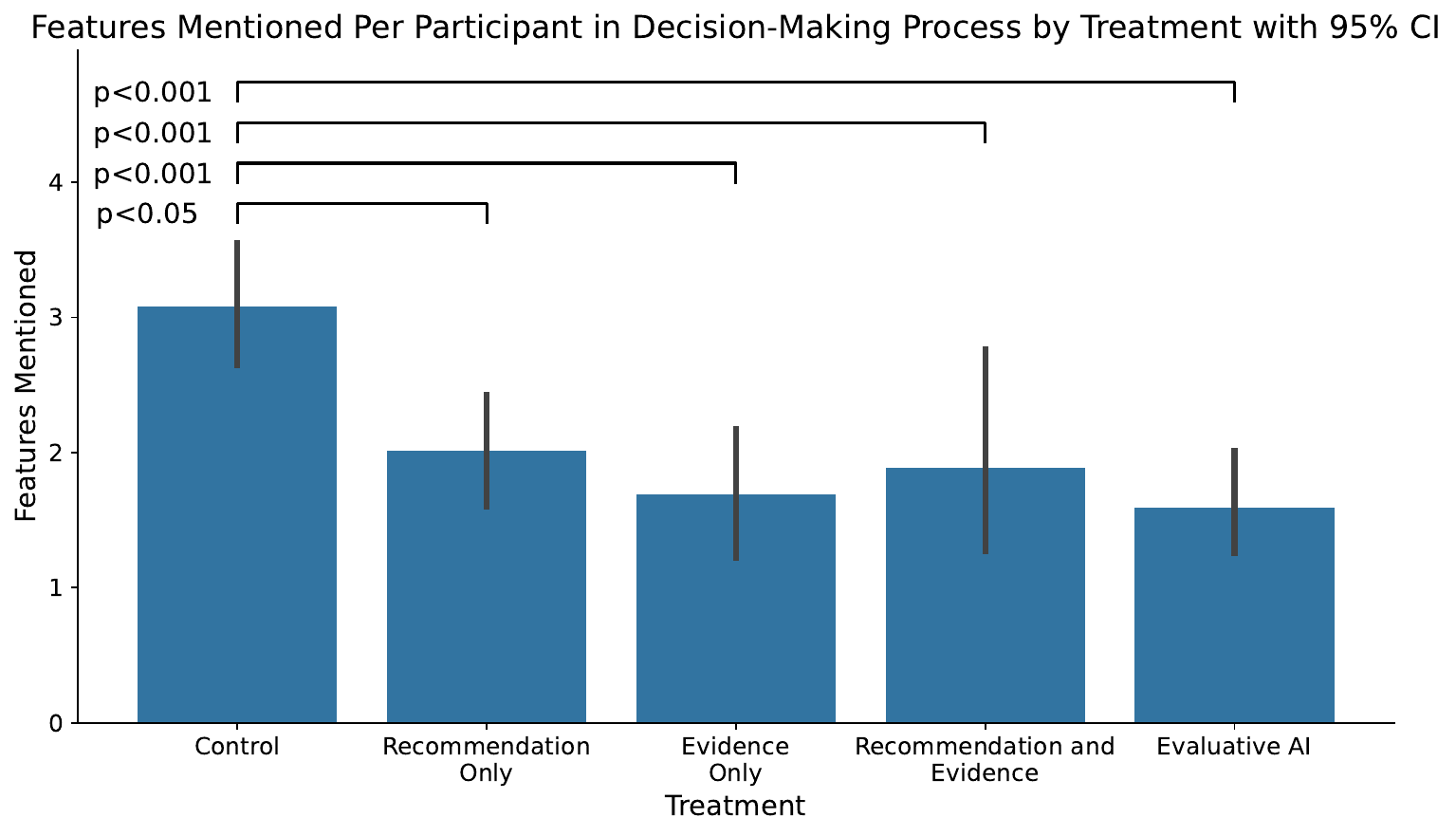}
    \captionsetup{width=0.9\linewidth}
    \caption{Average Number of Features Mentioned Per Participant by Treatment Group with 95\% Confidence Intervals. The bar chart depicts the mean number of features mentioned by participants during the decision-making process across five treatment groups: \co, \ro, \eo, \rae, and \eva. Significant differences between the Control group and other treatments are marked with p-values ($p<0.05$, $p<0.001$).}
    \label{fig:features_sum}
\end{figure*}

The participants mostly talked about which features they focused on for their assessment, and this differs significantly between \co and the other groups (pairwise chi-squared test, always $p < 0.001$). While in \co, 91.93\% of the participants mentioned at least one feature, the percentages were 66.23\% in \ro, 59.38\% in \eo, 50.62\% in \rae, and 54.95\% in \eva. An analysis of the number of mentioned features shows a similar pattern. On average, participants in \co mentioned 3.08 features, compared to 2.01 in \ro, 1.69 in \eo, 1.89 in \rae, and 1.59 in \eva. The differences between the groups with AI and \co are also significant according to the chi-squared test (with \ro, $p < 0.05$; otherwise, $p < 0.001$). \autoref{fig:features_proportion_mentioned} and \autoref{fig:features_sum} illustrate the proportion of participants who mentioned features and the average number of features used.

\begin{figure*}[]
    \centering
    \includegraphics[width=0.7\linewidth]{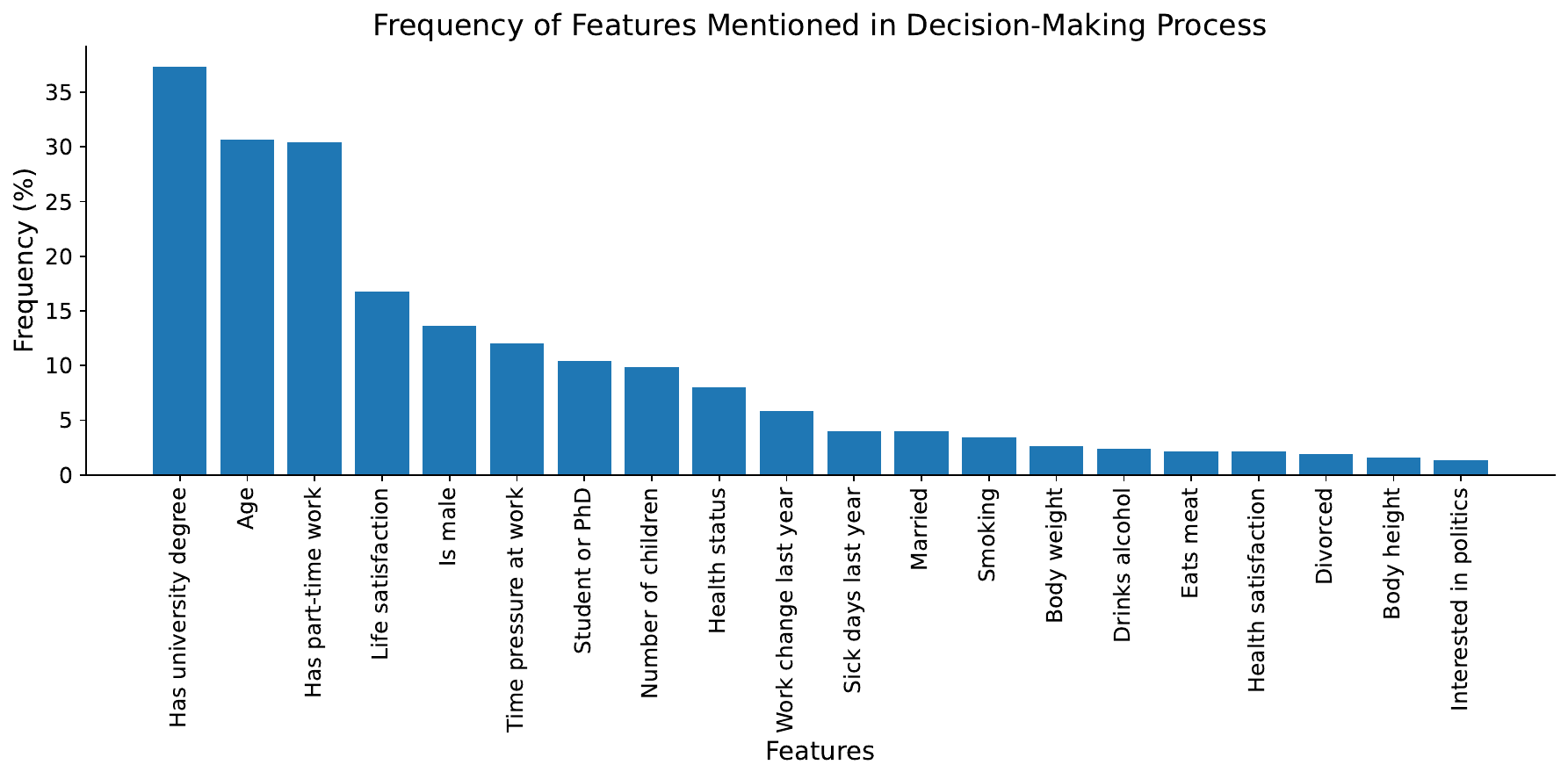}
    \captionsetup{width=0.9\linewidth}
    \caption{Frequency of Features Mentioned in the Decision-Making Process. The horizontal bar chart shows the percentage of participants mentioning specific features during the decision-making process.}
    \label{fig:features_frequency}
\end{figure*}

\begin{figure*}[]
    \centering
    \includegraphics[width=0.7\linewidth]{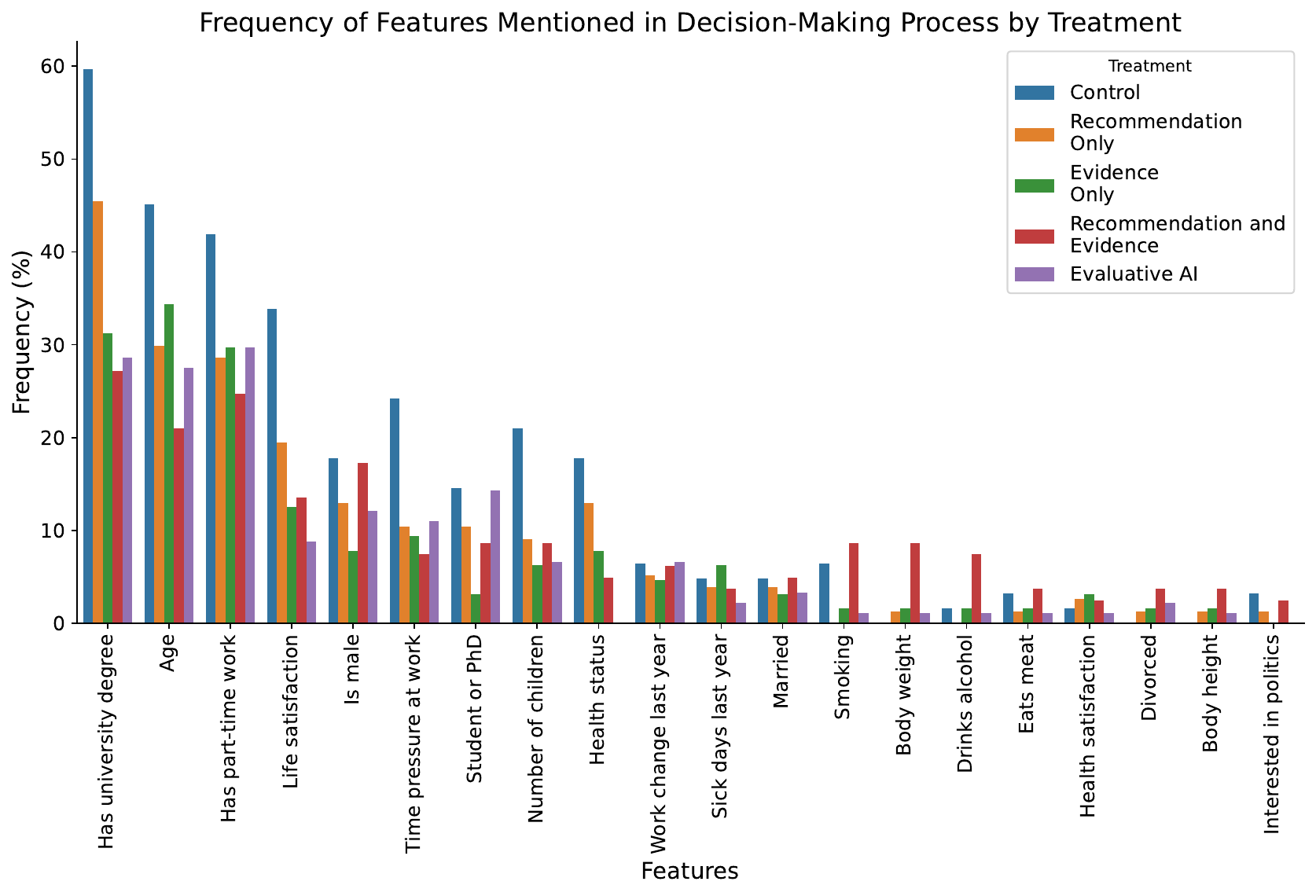}
    \captionsetup{width=0.9\linewidth}
    \caption{Frequency of Features Mentioned in the Decision-Making Process by Treatment. The horizontal bar chart shows the percentage of participants mentioning specific features during the decision-making process divided by treatments.}
    \label{fig:features_frequency_by_treatment}
\end{figure*}

\autoref{fig:features_frequency} shows the frequency of each feature mentioned in the participants' descriptions. Over 30\% of the descriptions included the features \textit{Has university degree}, \textit{Age}, and \textit{Has part-time work}. The fourth most mentioned feature was \textit{Life satisfaction}, at 16.8\%, after which the frequency steadily declines. The distribution per treatment in \autoref{fig:features_frequency_by_treatment} confirms the observation that features were mentioned more frequently in \co; however, there are no major differences between the features and the treatments.

\subsection{Usage of \eva}
In contrast to \eo, participants in \eva were not shown the pro and con evidence directly; instead, they had the freedom to display them at any time using buttons. The button clicks were tracked to analyze usage behavior.

\begin{figure*}
    \centering
    \includegraphics[width=0.6\linewidth]{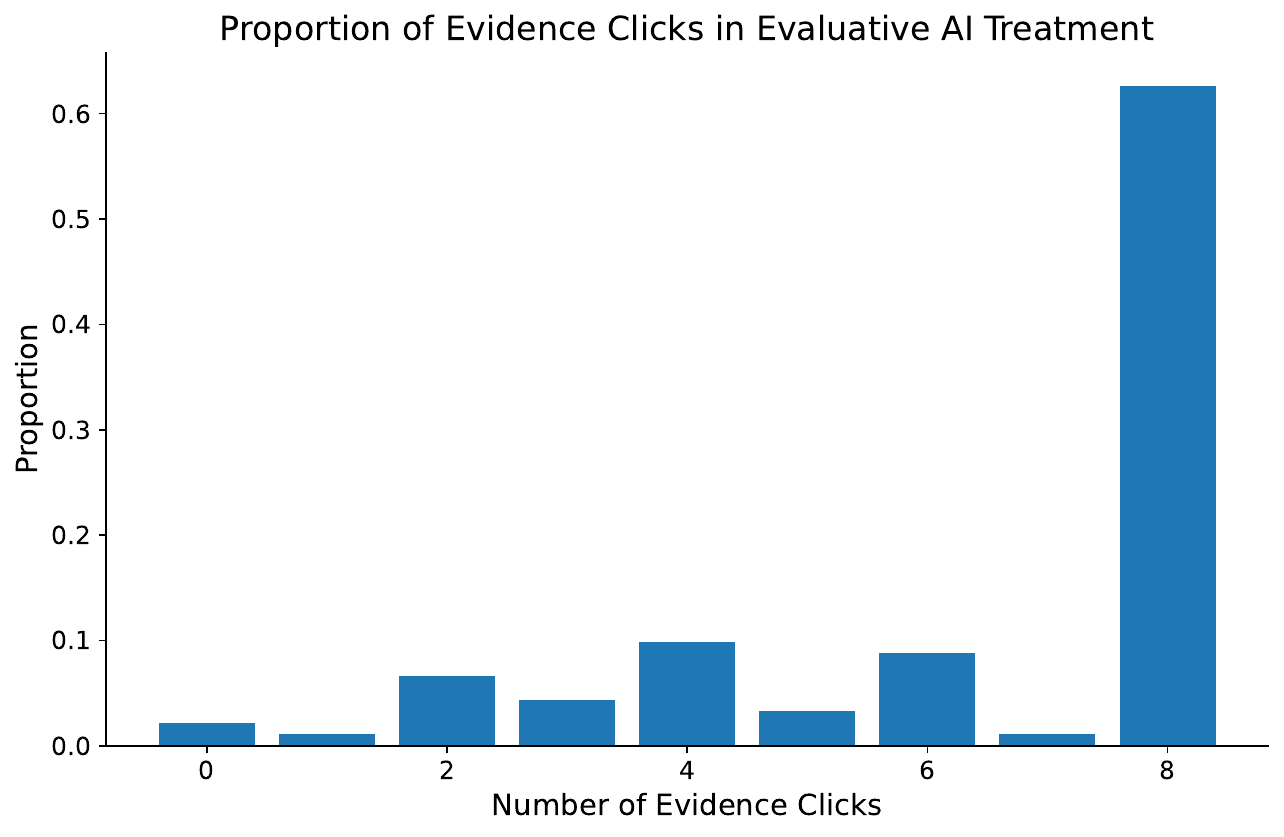}
    \captionsetup{width=0.9\linewidth}
    \caption{Distribution of Evidence Clicks in \eva. The bar chart illustrates the proportion of participants who clicked on varying numbers of evidence items (ranging from 0 to 8) during \eva.}
    \label{fig:number_evidence_click}
\end{figure*}

Of the 91 participants in \eva, 57 (62.64\%) clicked on the evidence in every round to display it. The remaining participants were relatively evenly distributed in terms of the number of clicks during the task. \autoref{fig:number_evidence_click} shows the distribution of clicks.

\begin{figure*}[h]
    \centering
    \includegraphics[width=0.6\linewidth]{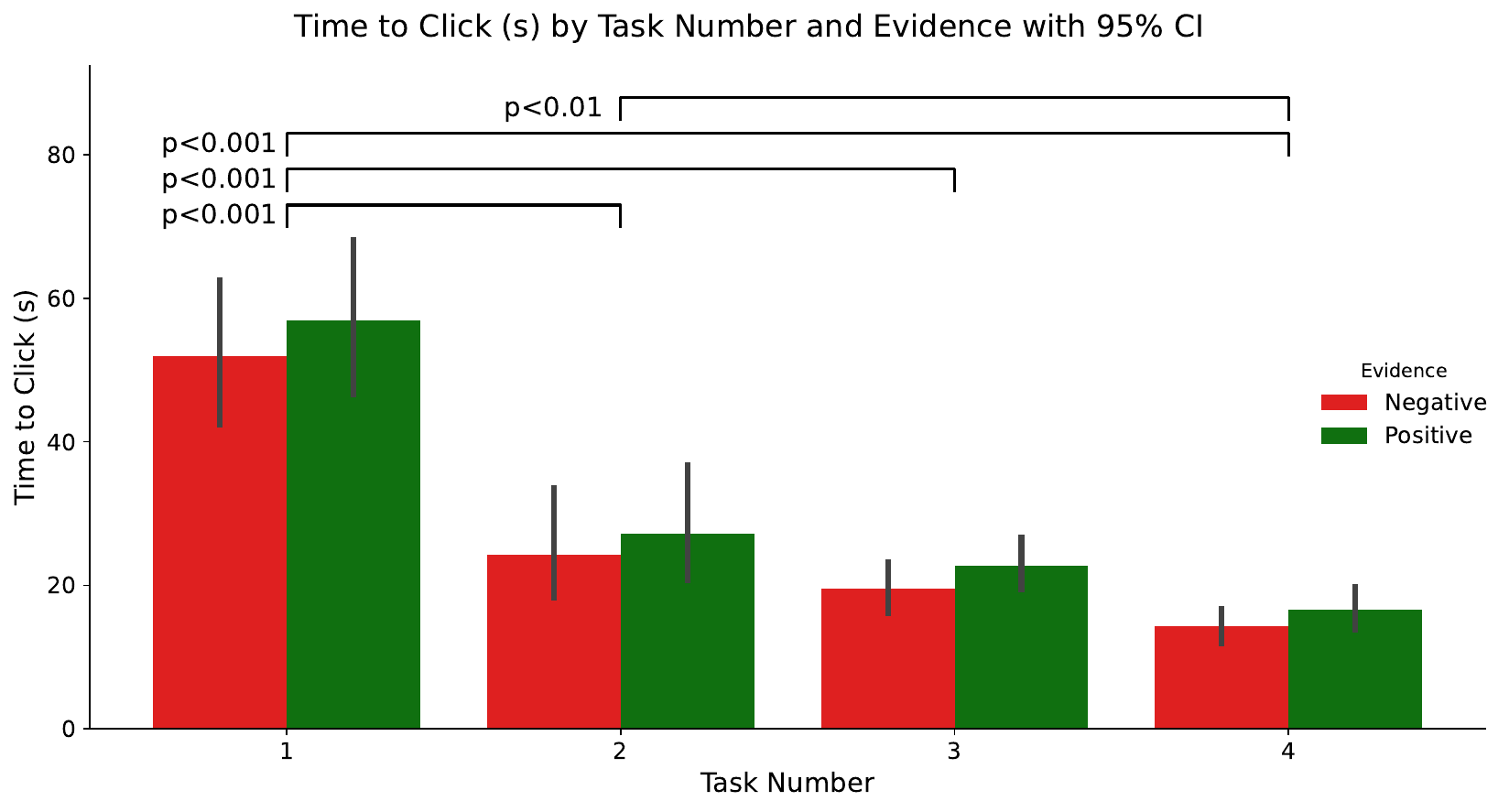}
    \captionsetup{width=0.9\linewidth}
    \caption{Time to Click (s) by Task Number and Evidence Type with 95\% Confidence Intervals. This bar chart shows the average time taken (in seconds) to click on negative evidence (red) or positive evidence (green), across four tasks. Significant differences between evidence types and task numbers are highlighted with p-values ($p<0.01$, $p<0.001)$.}
    \label{fig:time_to_click_by_task_and_evidence}
\end{figure*}

An examination of individual participants shows that, in most cases, they clicked on both pieces of evidence within a few seconds of each other. Figure \autoref{fig:time_to_click_by_task_and_evidence} illustrates the average time in seconds that participants in \eva took to view the evidence, broken down by the four tasks and the two types of evidence. It was also observed that participants took more time to click on the evidence during the first of the four tasks compared to the remaining tasks (Kruskal-Wallis test, $p < 0.001$).

\section{Discussion}
The aim of the present study is the empirical evaluation of the ``Evaluative AI'' framework proposed by \cite{Miller_2023}, specifically focusing on the assessment of pro and con evidence elements, which contrasts with traditional recommender-driven AI systems. The results of the behavioral experiment differed from the hypotheses: the AI based on the ``Evaluative AI'' framework did not improve participants' decision-making performance compared to treatments without AI assistance or with other types of AI support. Decision-making speed was also not the slowest, but it was significantly slower than in the control group and the group that received only AI recommendations. Cognitive load was not higher; there were no differences between the groups in this respect. The qualitative analysis of decision-making processes shows that the AI was similarly relevant for participants across the AI groups. Interestingly, participants often focused on the available features, and it was found that those without AI assistance discussed these features significantly more than participants in the other groups.

\textbf{Performance.} The most striking results concern performance. The fact that 73.44\% of participants performed better than random guessing suggests that they had some relevant knowledge and made an effort in completing the tasks. Unlike many studies that demonstrate AI recommendations can improve performance \citep{Hemmer_Schemmer_Kühl_Vössing_Satzger_2024,Hemmer_Schemmer_Vössing_Kühl_2021,Malone_Vaccaro_Campero_Song_Wen_Almaatouq_2023}, especially \cite{Le_Miller_Sonenberg_Singh_2024} conducting a similar evaluation, there was no significant improvement compared to the control group without AI assistance. One possible explanation could be that the AI was not significantly better than the participants.

The underlying ML model, however, is on par in quality (with an accuracy of 75\%) with models used in similar studies: \cite{Bucinca_Malaya_Gajos_2021} and \cite{Zhang_Liao_Bellamy_2020} also report 75\% accuracy, \cite{Wang_Yin_2021} 69\%, \cite{Bansal_Wu2021} 75–87\%, \cite{Liu_Lai_Tan_2021} 56–84\%, and \cite{Lai_Tan_2019} 87\%. In the control group, 43.55\% of participants outperformed the AI, while in the recommendation-only group, 55.84\% did so, suggesting the potential for complementary human-AI teamwork \citep{Hemmer_Schemmer_Kühl_Vössing_Satzger_2024}. 

On one hand, a bad performing AI could explain the lack of significant improvement. On the other hand, observations from other studies indicate that even when AI outperforms the control group by up to 15.5 percentage points, participants with AI assistance do not necessarily show better results \citep{Goh_Gallo_Hom_Strong_Weng_Kerman_Cool_Kanjee_Parsons_Ahuja_et_al._2024}. This might stem from algorithmic aversion  \citep{Mahmud_Islam_Ahmed_Smolander_2022, Castelo_Bos_Lehmann_2019}, though this explanation is inconsistent with the qualitative results, as many participants considered the AI’s input.

Even though it may seem disappointing from the perspective of the ``Evaluative AI'' framework that performance did not improve, this result aligns with the mixed findings in XAI research. While the framework itself does not directly focus on explanations but rather on the overall decision-making process, studies show that the effects of explanations are not conclusive. For instance, while \cite{Lai_Tan_2019} and \cite{Lai_Liu_Tan_2020} found that explanations (with and without recommendations) positively impacted performance, there are also opposing findings: \cite{Bansal_Wu2021} reported increased performance due to AI recommendations, but no further improvement from explanations, and \cite{Zhang_Liao_Bellamy_2020} similarly found no effect from XAI. One reason could be the SHAP explanations used;  \cite{Kaur_Nori_Jenkins_Caruana_Wallach_Wortman_Vaughan_2020} found that even data scientists struggled with bar chart-like tools. To counter this, textual explanations were also provided in the present study.

\textbf{Decision Time and Cognitive Load.} The fact that participants noticed the explanations is evident in the analysis of processing speed: all groups with explanations were slower than both the control group and the recommendation-only group. \cite{Carton_Mei_Resnick_2020} reported an increased decision time due to recommendations, but a simultaneous reduction with an explanation for the recommendation. \cite{Cheng_Wang_Zhang_O’Connell_Gray_Harper_Zhu_2019} and \cite{Slack_Friedler_Scheidegger_Roy_2019} found that increased transparency costs more time. 

Despite differences in decision time, however, there were no significant differences in subjectively measured cognitive load, contradicting findings by \cite{Herm_2023}, who observed a linear relationship between task time and cognitive load. One reason for the differing result in this study could be that, although there were significant time differences between treatments with and without explanations, the differences were small (about 17 seconds on average between \co and \rae). This may not be sufficient to place a greater cognitive demand on participants, especially given that there were only four tasks in total, so the overall time difference was minimal.

\textbf{Decision-Making Process and Engagement.} The qualitative analysis of the decision-making processes reveals that participants engaged cognitively with recommendations and weighed pro and con evidence. First, between 37 and 53\% of participants across various treatments mentioned the AI in their descriptions. More importantly, participants who had AI support relied significantly less on specific features in their descriptions. This suggests cognitive offloading \citep{Risko_Gilbert_2016} may have occurred, along with a potential automation bias \citep{Lyell_Coiera_2017}. Automation bias leads to uncalibrated use of AI, often resulting in overreliance. Reducing overreliance is one of the key motivations behind the ``Evaluative AI'' framework. Nonetheless, participants' cognitive processes in \eva did not appear markedly different from those in other treatments.

One reason for this could be that not all users engaged with the pro and con evidence. 62.6\% of participants reviewed both sides of the evidence in all rounds. This pattern of superficial engagement with explanations is not new \citep{Bucinca_Malaya_Gajos_2021}. The lack of interest in provided evidence among some participants could be due to a degree of algorithm aversion. Even though instructions explained the AI’s performance, participants did not experience it personally and may therefore have lacked trust. Participants may also have made a cost-benefit assessment; according to \cite{Vasconcelos_2023}, participants evaluate whether engaging with provided evidence is worth their time. Although this study attempted to create a high-stakes environment with substantial task-based bonuses, these incentives may not have been high enough to motivate participants toward deeper engagement.

\section{Limitations and Future Work}
Although the ``Evaluative AI'' framework is theoretically well-founded, with \cite{Le_Miller_Sonenberg_Singh_2024} reporting promising results in similar studies, the present study reveals that implementing and examining such a framework in practice is challenging. There are several points future researchers and practitioners should consider.

Contrary to expectations, no performance improvements could be measured using an AI system based on the framework. One aspect worth discussing is the fundamental machine learning model used, along with the generated evidence. The model applied here did not significantly outperform the participants, which may have contributed to the absence of notable improvements. Nevertheless, it was comparable to models from related literature. Even though \cite{Goh_Gallo_Hom_Strong_Weng_Kerman_Cool_Kanjee_Parsons_Ahuja_et_al._2024} noted that improvements are not guaranteed under these circumstances, this comparison may be an essential baseline to achieve.

The pro and con evidence should be presented in a way that is clear and accessible to users. This study found that many participants did not make use of them. While XAI research offers various options for optimally presenting explanations, research specifically focusing on hypothesis-driven AI could investigate ways to improve the clarity and usability of these presentations. Mixed-methods approaches should also be applied to better understand participants' decision-making processes.

Another relevant point is the importance of testing AI systems across enough domains to ensure external validity. Previous research has shown multiple times that results can be influenced by the domains in which they are applied \citep{Le_Miller_Singh_Sonenberg_2023,Kornowicz_Thommes_2024,Bogard_Shu_2022}. 

One further limitation is the use of laypeople for empirical evaluation. \cite{Miller_2023} argued that the framework should ideally be applied in medium/high-stakes situations, which likely require domain-specific knowledge. Lastly, the decision problem could be expanded from binary to multi-class decisions. For example, \cite{Miller_2023} presents a diagnostic scenario involving multiple diseases, where several hypotheses can be individually assessed.

\section{Conclusion}
The present study examines the effectiveness of the ``Evaluative AI'' framework, focusing on the provision of pro and con evidence within a hypothesis-driven AI approach. Results from the behavioral experiment paint a sobering picture: decision-making performance did not improve; instead, all participants who received evidence from the AI were slower in making decisions, although cognitive load remained unaffected. Qualitative data indicated that all AI systems led to a form of cognitive offloading and potential automation bias, with a significant portion of participants engaging only superficially with the evidence presented.

Although the study questions the empirical validity of the proposed framework, there are limitations that should be addressed in future research. These include developing appropriate AI systems, investigating the presentation of pro and con evidence, considering alternative forms of decision-making, involving domain-specific experiments, and better simulating high-stakes situations. Despite the present findings, the evaluative AI framework is a well-conceived model with the potential to be a promising direction for AI-based decision support.

\section*{Acknowledgements}
The author(s) would like to thank Kirsten Thommes for her detailed feedback. 

\section*{Funding} The author gratefully acknowledges funding by the German Research Foundation (Deutsche
Forschungsgemeinschaft, DFG): TRR 318/1 2021 – 438445824

\section*{Disclosure statement}
The author(s) report there are no competing interests to declare.

\section*{Data availability statement}
The code for the program software, the experiment data, and the analysis code can be found in the \href{https://osf.io/7pbt2/?view_only=12447fb8850a4ba68628df16b4536a9b}{public repository}.

\section*{Institutional review board statement} 
The ethics board of the University of Paderborn approved the research project.

\end{multicols}

\bibliography{references}

\begin{thebibliography}{}

\bibitem[\protect\astroncite{Albrecht}{2016}]{albrecht2016gdpr}
Albrecht, J.~P. (2016).
\newblock How the gdpr will change the world.
\newblock {\em Eur. Data Prot. L. Rev.}, 2:287.

\bibitem[\protect\astroncite{Bansal et~al.}{2021}]{Bansal_Wu2021}
Bansal, G., Wu, T., Zhou, J., Fok, R., Nushi, B., Kamar, E., Ribeiro, M.~T.,
  and Weld, D.~S. (2021).
\newblock Does the whole exceed its parts? the effect of ai explanations on
  complementary team performance.
\newblock (arXiv:2006.14779).
\newblock arXiv:2006.14779 [cs].

\bibitem[\protect\astroncite{Barredo~Arrieta et~al.}{2020}]{Barredo_2020}
Barredo~Arrieta, A., Díaz-Rodríguez, N., Del~Ser, J., Bennetot, A., Tabik,
  S., Barbado, A., Garcia, S., Gil-Lopez, S., Molina, D., Benjamins, R.,
  Chatila, R., and Herrera, F. (2020).
\newblock Explainable artificial intelligence (xai): Concepts, taxonomies,
  opportunities and challenges toward responsible ai.
\newblock {\em Information Fusion}, 58:82–115.

\bibitem[\protect\astroncite{Becker and Kohavi}{1996}]{adult_2}
Becker, B. and Kohavi, R. (1996).
\newblock {Adult}.
\newblock UCI Machine Learning Repository.
\newblock {DOI}: https://doi.org/10.24432/C5XW20.

\bibitem[\protect\astroncite{Bertrand
  et~al.}{2023}]{Bertrand_Viard_Belloum_Eagan_Maxwell_2023}
Bertrand, A., Viard, T., Belloum, R., Eagan, J.~R., and Maxwell, W. (2023).
\newblock On selective, mutable and dialogic xai: a review of what users say
  about different types of interactive explanations.
\newblock In {\em Proceedings of the 2023 CHI Conference on Human Factors in
  Computing Systems}, page 1–21, Hamburg Germany. ACM.

\bibitem[\protect\astroncite{Bogard and Shu}{2022}]{Bogard_Shu_2022}
Bogard, J. and Shu, S. (2022).
\newblock {\em Algorithm Aversion and the Aversion to Counter-Normative
  Decision Procedures}.

\bibitem[\protect\astroncite{Buçinca et~al.}{2021}]{Bucinca_Malaya_Gajos_2021}
Buçinca, Z., Malaya, M.~B., and Gajos, K.~Z. (2021).
\newblock To trust or to think: Cognitive forcing functions can reduce
  overreliance on ai in ai-assisted decision-making.
\newblock {\em Proceedings of the ACM on Human-Computer Interaction},
  5(CSCW1):188:1--188:21.

\bibitem[\protect\astroncite{Buçinca et~al.}{2024}]{Bucinca__2024_contrastive}
Buçinca, Z., Swaroop, S., Paluch, A.~E., Doshi-Velez, F., and Gajos, K.~Z.
  (2024).
\newblock Contrastive explanations that anticipate human misconceptions can
  improve human decision-making skills.
\newblock (arXiv:2410.04253).
\newblock arXiv:2410.04253.

\bibitem[\protect\astroncite{Carton et~al.}{2020}]{Carton_Mei_Resnick_2020}
Carton, S., Mei, Q., and Resnick, P. (2020).
\newblock Feature-based explanations don’t help people detect
  misclassifications of online toxicity.
\newblock {\em Proceedings of the International AAAI Conference on Web and
  Social Media}, 14:95–106.

\bibitem[\protect\astroncite{Castelnovo
  et~al.}{2023}]{Castelnovo_Crupi_Mombelli_Nanino_Regoli_2023}
Castelnovo, A., Crupi, R., Mombelli, N., Nanino, G., and Regoli, D. (2023).
\newblock Evaluative item-contrastive explanations in rankings.
\newblock (arXiv:2312.10094).
\newblock arXiv:2312.10094 [cs].

\bibitem[\protect\astroncite{Castelo et~al.}{2019}]{Castelo_Bos_Lehmann_2019}
Castelo, N., Bos, M.~W., and Lehmann, D.~R. (2019).
\newblock Task-dependent algorithm aversion.
\newblock {\em Journal of Marketing Research}, 56(5):809–825.

\bibitem[\protect\astroncite{Chen et~al.}{2016}]{chen2016otree}
Chen, D.~L., Schonger, M., and Wickens, C. (2016).
\newblock otree—an open-source platform for laboratory, online, and field
  experiments.
\newblock {\em Journal of Behavioral and Experimental Finance}, 9:88--97.

\bibitem[\protect\astroncite{Chen et~al.}{2023}]{Chen_Liao_Vaughan_Bansal_2023}
Chen, V., Liao, Q.~V., Vaughan, J.~W., and Bansal, G. (2023).
\newblock Understanding the role of human intuition on reliance in human-ai
  decision-making with explanations.
\newblock (arXiv:2301.07255).
\newblock arXiv:2301.07255 [cs].

\bibitem[\protect\astroncite{Cheng
  et~al.}{2019}]{Cheng_Wang_Zhang_O’Connell_Gray_Harper_Zhu_2019}
Cheng, H.-F., Wang, R., Zhang, Z., O’Connell, F., Gray, T., Harper, F.~M.,
  and Zhu, H. (2019).
\newblock Explaining decision-making algorithms through ui: Strategies to help
  non-expert stakeholders.
\newblock In {\em Proceedings of the 2019 CHI Conference on Human Factors in
  Computing Systems}, CHI ’19, page 1–12, New York, NY, USA. Association
  for Computing Machinery.

\bibitem[\protect\astroncite{Chew
  et~al.}{2023}]{Chew_Bollenbacher_Wenger_Speer_Kim_2023}
Chew, R., Bollenbacher, J., Wenger, M., Speer, J., and Kim, A. (2023).
\newblock Llm-assisted content analysis: Using large language models to support
  deductive coding.
\newblock (arXiv:2306.14924).
\newblock arXiv:2306.14924.

\bibitem[\protect\astroncite{Chromik
  et~al.}{2021}]{Chromik_Eiband_Buchner_Krüger_Butz_2021}
Chromik, M., Eiband, M., Buchner, F., Krüger, A., and Butz, A. (2021).
\newblock I think i get your point, ai! the illusion of explanatory depth in
  explainable ai.
\newblock In {\em 26th International Conference on Intelligent User
  Interfaces}, IUI ’21, page 307–317, New York, NY, USA. Association for
  Computing Machinery.

\bibitem[\protect\astroncite{Cresswell
  et~al.}{2024}]{Cresswell_Sui_Kumar_Vouitsis_2024}
Cresswell, J.~C., Sui, Y., Kumar, B., and Vouitsis, N. (2024).
\newblock Conformal prediction sets improve human decision making.
\newblock (arXiv:2401.13744).
\newblock arXiv:2401.13744 [cs, stat].

\bibitem[\protect\astroncite{Croskerry}{2009}]{croskerry2009universal}
Croskerry, P. (2009).
\newblock A universal model of diagnostic reasoning.
\newblock {\em Academic medicine}, 84(8):1022--1028.

\bibitem[\protect\astroncite{Faul et~al.}{2007}]{gpower}
Faul, F., Erdfelder, E., Lang, A.-G., and Buchner, A. (2007).
\newblock G* power 3: A flexible statistical power analysis program for the
  social, behavioral, and biomedical sciences.
\newblock {\em Behavior research methods}, 39(2):175--191.

\bibitem[\protect\astroncite{Gajos and Mamykina}{2022}]{Gajos_Mamykina_2022}
Gajos, K.~Z. and Mamykina, L. (2022).
\newblock Do people engage cognitively with ai? impact of ai assistance on
  incidental learning.
\newblock In {\em 27th International Conference on Intelligent User
  Interfaces}, page 794–806, Helsinki Finland. ACM.

\bibitem[\protect\astroncite{Ghai et~al.}{2020}]{Ghai2020}
Ghai, B., Liao, Q.~V., Zhang, Y., Bellamy, R., and Mueller, K. (2020).
\newblock Explainable active learning (xal): An empirical study of how local
  explanations impact annotator experience.
\newblock (arXiv:2001.09219).
\newblock arXiv:2001.09219 [cs].

\bibitem[\protect\astroncite{Goebel et~al.}{2019}]{goebel2019german}
Goebel, J., Grabka, M.~M., Liebig, S., Kroh, M., Richter, D., Schr{\"o}der, C.,
  and Schupp, J. (2019).
\newblock The german socio-economic panel (soep).
\newblock {\em Jahrb{\"u}cher f{\"u}r National{\"o}konomie und Statistik},
  239(2):345--360.

\bibitem[\protect\astroncite{Goh
  et~al.}{2024}]{Goh_Gallo_Hom_Strong_Weng_Kerman_Cool_Kanjee_Parsons_Ahuja_et_al._2024}
Goh, E., Gallo, R., Hom, J., Strong, E., Weng, Y., Kerman, H., Cool, J.,
  Kanjee, Z., Parsons, A.~S., Ahuja, N., Horvitz, E., Yang, D., Milstein, A.,
  Olson, A.~P., Rodman, A., and Chen, J.~H. (2024).
\newblock Influence of a large language model on diagnostic reasoning: A
  randomized clinical vignette study.
\newblock {\em medRxiv}, page 2024.03.12.24303785.

\bibitem[\protect\astroncite{Gouveia and Malík}{2024}]{Gouveia_2024}
Gouveia, S.~S. and Malík, J. (2024).
\newblock Crossing the trust gap in medical ai: Building an abductive bridge
  for xai.
\newblock {\em Philosophy \& Technology}, 37(3):105.

\bibitem[\protect\astroncite{Guidotti et~al.}{2018}]{Guidotti_2018}
Guidotti, R., Monreale, A., Ruggieri, S., Turini, F., Giannotti, F., and
  Pedreschi, D. (2018).
\newblock A survey of methods for explaining black box models.
\newblock {\em ACM Computing Surveys}, 51(5):93:1--93:42.

\bibitem[\protect\astroncite{Hart}{2006}]{Hart_2006}
Hart, S.~G. (2006).
\newblock Nasa-task load index (nasa-tlx); 20 years later.
\newblock {\em Proceedings of the Human Factors and Ergonomics Society Annual
  Meeting}, 50(9):904–908.

\bibitem[\protect\astroncite{Hemmer
  et~al.}{2024}]{Hemmer_Schemmer_Kühl_Vössing_Satzger_2024}
Hemmer, P., Schemmer, M., Kühl, N., Vössing, M., and Satzger, G. (2024).
\newblock Complementarity in human-ai collaboration: Concept, sources, and
  evidence.
\newblock (arXiv:2404.00029).
\newblock arXiv:2404.00029 [cs].

\bibitem[\protect\astroncite{Hemmer
  et~al.}{2021}]{Hemmer_Schemmer_Vössing_Kühl_2021}
Hemmer, P., Schemmer, M., Vössing, M., and Kühl, N. (2021).
\newblock Human-ai complementarity in hybrid intelligence systems: A structured
  literature review.
\newblock In {\em PACIS 2021 Proceedings}.

\bibitem[\protect\astroncite{Herm}{2023}]{Herm_2023}
Herm, L.-V. (2023).
\newblock Impact of explainable ai on cognitive load: Insights from an
  empirical study.

\bibitem[\protect\astroncite{Jacovi
  et~al.}{2021}]{Jacovi_Marasović_Miller_Goldberg_2021}
Jacovi, A., Marasović, A., Miller, T., and Goldberg, Y. (2021).
\newblock Formalizing trust in artificial intelligence: Prerequisites, causes
  and goals of human trust in ai.
\newblock In {\em Proceedings of the 2021 ACM Conference on Fairness,
  Accountability, and Transparency}, FAccT ’21, page 624–635, New York, NY,
  USA. Association for Computing Machinery.

\bibitem[\protect\astroncite{Kaur
  et~al.}{2020}]{Kaur_Nori_Jenkins_Caruana_Wallach_Wortman_Vaughan_2020}
Kaur, H., Nori, H., Jenkins, S., Caruana, R., Wallach, H., and Wortman~Vaughan,
  J. (2020).
\newblock Interpreting interpretability: Understanding data scientists’ use
  of interpretability tools for machine learning.
\newblock In {\em Proceedings of the 2020 CHI Conference on Human Factors in
  Computing Systems}, CHI ’20, page 1–14, New York, NY, USA. Association
  for Computing Machinery.

\bibitem[\protect\astroncite{Klein et~al.}{2007}]{klein2007data}
Klein, G., Phillips, J.~K., Rall, E.~L., and Peluso, D.~A. (2007).
\newblock A data--frame theory of sensemaking.
\newblock In {\em Expertise out of context}, pages 118--160. Psychology Press.

\bibitem[\protect\astroncite{Kornowicz and
  Thommes}{2024}]{Kornowicz_Thommes_2024}
Kornowicz, J. and Thommes, K. (2024).
\newblock Algorithm, expert, or both? evaluating the role of feature selection
  methods on user preferences and reliance.
\newblock (arXiv:2408.01171).
\newblock arXiv:2408.01171 [cs].

\bibitem[\protect\astroncite{Lai
  et~al.}{2023a}]{Lai_Chen_Smith-Renner_Liao_Tan_2023}
Lai, V., Chen, C., Smith-Renner, A., Liao, Q.~V., and Tan, C. (2023a).
\newblock Towards a science of human-ai decision making: An overview of design
  space in empirical human-subject studies.
\newblock In {\em 2023 ACM Conference on Fairness, Accountability, and
  Transparency}, page 1369–1385, Chicago IL USA. ACM.

\bibitem[\protect\astroncite{Lai et~al.}{2020}]{Lai_Liu_Tan_2020}
Lai, V., Liu, H., and Tan, C. (2020).
\newblock “why is ‘chicago’ deceptive?” towards building model-driven
  tutorials for humans.
\newblock In {\em Proceedings of the 2020 CHI Conference on Human Factors in
  Computing Systems}, page 1–13, Honolulu HI USA. ACM.

\bibitem[\protect\astroncite{Lai and Tan}{2019}]{Lai_Tan_2019}
Lai, V. and Tan, C. (2019).
\newblock On human predictions with explanations and predictions of machine
  learning models: A case study on deception detection.
\newblock In {\em Proceedings of the Conference on Fairness, Accountability,
  and Transparency}, page 29–38, Atlanta GA USA. ACM.

\bibitem[\protect\astroncite{Lai et~al.}{2023b}]{Lai_Zhang_Chen_Liao_Tan_2023}
Lai, V., Zhang, Y., Chen, C., Liao, Q.~V., and Tan, C. (2023b).
\newblock Selective explanations: Leveraging human input to align explainable
  ai.
\newblock (arXiv:2301.09656).
\newblock arXiv:2301.09656 [cs].

\bibitem[\protect\astroncite{Le et~al.}{2023}]{Le_Miller_Singh_Sonenberg_2023}
Le, T., Miller, T., Singh, R., and Sonenberg, L. (2023).
\newblock Explaining model confidence using counterfactuals.
\newblock {\em Proceedings of the AAAI Conference on Artificial Intelligence},
  37(1010):11856–11864.

\bibitem[\protect\astroncite{Le et~al.}{2024a}]{Le_Miller_Sonenberg_Singh_2024}
Le, T., Miller, T., Sonenberg, L., and Singh, R. (2024a).
\newblock Towards the new xai: A hypothesis-driven approach to decision support
  using evidence.
\newblock (arXiv:2402.01292).
\newblock arXiv:2402.01292 [cs].

\bibitem[\protect\astroncite{Le
  et~al.}{2024b}]{Le_Miller_Zhang_Sonenberg_Singh_2024}
Le, T., Miller, T., Zhang, R., Sonenberg, L., and Singh, R. (2024b).
\newblock Visual evaluative ai: A hypothesis-driven tool with concept-based
  explanations and weight of evidence.
\newblock (arXiv:2407.04710).
\newblock arXiv:2407.04710 [cs].

\bibitem[\protect\astroncite{Liu et~al.}{2021}]{Liu_Lai_Tan_2021}
Liu, H., Lai, V., and Tan, C. (2021).
\newblock Understanding the effect of out-of-distribution examples and
  interactive explanations on human-ai decision making.
\newblock {\em Proceedings of the ACM on Human-Computer Interaction},
  5(CSCW2):408:1--408:45.

\bibitem[\protect\astroncite{Lundberg and Lee}{2017}]{shap_paper}
Lundberg, S.~M. and Lee, S.-I. (2017).
\newblock A unified approach to interpreting model predictions.
\newblock In Guyon, I., Luxburg, U.~V., Bengio, S., Wallach, H., Fergus, R.,
  Vishwanathan, S., and Garnett, R., editors, {\em Advances in Neural
  Information Processing Systems 30}, pages 4765--4774. Curran Associates, Inc.

\bibitem[\protect\astroncite{Lyell and Coiera}{2017}]{Lyell_Coiera_2017}
Lyell, D. and Coiera, E. (2017).
\newblock Automation bias and verification complexity: a systematic review.
\newblock {\em Journal of the American Medical Informatics Association},
  24(2):423–431.

\bibitem[\protect\astroncite{Ma
  et~al.}{2023}]{Ma_Lei_Wang_Zheng_Shi_Yin_Ma_2023}
Ma, S., Lei, Y., Wang, X., Zheng, C., Shi, C., Yin, M., and Ma, X. (2023).
\newblock Who should i trust: Ai or myself? leveraging human and ai correctness
  likelihood to promote appropriate trust in ai-assisted decision-making.
\newblock In {\em Proceedings of the 2023 CHI Conference on Human Factors in
  Computing Systems}, page 1–19, Hamburg Germany. ACM.

\bibitem[\protect\astroncite{MacCarthy}{2019}]{maccarthy2019examination}
MacCarthy, M. (2019).
\newblock An examination of the algorithmic accountability act of 2019.
\newblock {\em Available at SSRN 3615731}.

\bibitem[\protect\astroncite{Mahmud
  et~al.}{2022}]{Mahmud_Islam_Ahmed_Smolander_2022}
Mahmud, H., Islam, A. K. M.~N., Ahmed, S.~I., and Smolander, K. (2022).
\newblock What influences algorithmic decision-making? a systematic literature
  review on algorithm aversion.
\newblock {\em Technological Forecasting and Social Change}, 175:121390.

\bibitem[\protect\astroncite{Malone
  et~al.}{2023}]{Malone_Vaccaro_Campero_Song_Wen_Almaatouq_2023}
Malone, T., Vaccaro, M., Campero, A., Song, J., Wen, H., and Almaatouq, A.
  (2023).
\newblock A test for evaluating performance in human-ai systems.

\bibitem[\protect\astroncite{Mayring}{2015}]{Mayring_2015}
Mayring, P. (2015).
\newblock {\em Qualitative Content Analysis: Theoretical Background and
  Procedures}, page 365–380.
\newblock Springer Netherlands, Dordrecht.

\bibitem[\protect\astroncite{Miller}{2019}]{Miller_2019}
Miller, T. (2019).
\newblock Explanation in artificial intelligence: Insights from the social
  sciences.
\newblock {\em Artificial Intelligence}, 267:1–38.

\bibitem[\protect\astroncite{Miller}{2023}]{Miller_2023}
Miller, T. (2023).
\newblock Explainable ai is dead, long live explainable ai!: Hypothesis-driven
  decision support using evaluative ai.
\newblock In {\em 2023 ACM Conference on Fairness, Accountability, and
  Transparency}, page 333–342, Chicago IL USA. ACM.

\bibitem[\protect\astroncite{Mnih et~al.}{2015}]{Mnih_2015}
Mnih, V., Kavukcuoglu, K., Silver, D., Rusu, A.~A., Veness, J., Bellemare,
  M.~G., Graves, A., Riedmiller, M., Fidjeland, A.~K., Ostrovski, G., Petersen,
  S., Beattie, C., Sadik, A., Antonoglou, I., King, H., Kumaran, D., Wierstra,
  D., Legg, S., and Hassabis, D. (2015).
\newblock Human-level control through deep reinforcement learning.
\newblock {\em Nature}, 518(75407540):529–533.

\bibitem[\protect\astroncite{Nori et~al.}{2023}]{nori2023capabilities}
Nori, H., King, N., McKinney, S.~M., Carignan, D., and Horvitz, E. (2023).
\newblock Capabilities of gpt-4 on medical challenge problems.
\newblock {\em arXiv preprint arXiv:2303.13375}.

\bibitem[\protect\astroncite{Peirce}{2009}]{peirce2009writings}
Peirce, C.~S. (2009).
\newblock {\em Writings of Charles S. Peirce: a chronological edition, volume
  8: 1890--1892}, volume~8.
\newblock Indiana University Press.

\bibitem[\protect\astroncite{Popper}{2014}]{popper2014conjectures}
Popper, K. (2014).
\newblock {\em Conjectures and refutations: The growth of scientific
  knowledge}.
\newblock routledge.

\bibitem[\protect\astroncite{Poursabzi-Sangdeh et~al.}{2021}]{Poursabzi2021}
Poursabzi-Sangdeh, F., Goldstein, D.~G., Hofman, J.~M., Wortman~Vaughan, J.~W.,
  and Wallach, H. (2021).
\newblock Manipulating and measuring model interpretability.
\newblock In {\em Proceedings of the 2021 CHI Conference on Human Factors in
  Computing Systems}, CHI ’21, page 1–52, New York, NY, USA. Association
  for Computing Machinery.

\bibitem[\protect\astroncite{Ribeiro
  et~al.}{2018}]{Ribeiro_Singh_Guestrin_2018}
Ribeiro, M.~T., Singh, S., and Guestrin, C. (2018).
\newblock Anchors: High-precision model-agnostic explanations.
\newblock {\em Proceedings of the AAAI Conference on Artificial Intelligence},
  32(11).

\bibitem[\protect\astroncite{Risko and Gilbert}{2016}]{Risko_Gilbert_2016}
Risko, E.~F. and Gilbert, S.~J. (2016).
\newblock Cognitive offloading.
\newblock {\em Trends in Cognitive Sciences}, 20(9):676–688.

\bibitem[\protect\astroncite{Rogha}{2023}]{Rogha_2023}
Rogha, M. (2023).
\newblock Explain to decide: A human-centric review on the role of explainable
  artificial intelligence in ai-assisted decision making.
\newblock (arXiv:2312.11507).
\newblock arXiv:2312.11507 [cs].

\bibitem[\protect\astroncite{Rong et~al.}{2022}]{Rong_2022}
Rong, Y., Leemann, T., Nguyen, T.-t., Fiedler, L., Seidel, T., Kasneci, G., and
  Kasneci, E. (2022).
\newblock Towards human-centered explainable ai: User studies for model
  explanations.
\newblock (arXiv:2210.11584).
\newblock arXiv:2210.11584 [cs].

\bibitem[\protect\astroncite{Rudin}{2019}]{Rudin_2019}
Rudin, C. (2019).
\newblock Stop explaining black box machine learning models for high stakes
  decisions and use interpretable models instead.
\newblock {\em Nature Machine Intelligence}, 1(5):206–215.

\bibitem[\protect\astroncite{Sawyer}{2009}]{Sawyer_2009}
Sawyer, S.~F. (2009).
\newblock Analysis of variance: The fundamental concepts.
\newblock {\em Journal of Manual \& Manipulative Therapy}.

\bibitem[\protect\astroncite{Schemmer et~al.}{2022}]{Schemmer_2022}
Schemmer, M., Hemmer, P., Nitsche, M., Kühl, N., and Vössing, M. (2022).
\newblock A meta-analysis of the utility of explainable artificial intelligence
  in human-ai decision-making.
\newblock In {\em Proceedings of the 2022 AAAI/ACM Conference on AI, Ethics,
  and Society}, page 617–626.
\newblock arXiv:2205.05126 [cs].

\bibitem[\protect\astroncite{Schemmer
  et~al.}{2023}]{Schemmer_Kuehl_Benz_Bartos_Satzger_2023}
Schemmer, M., Kuehl, N., Benz, C., Bartos, A., and Satzger, G. (2023).
\newblock Appropriate reliance on ai advice: Conceptualization and the effect
  of explanations.
\newblock In {\em Proceedings of the 28th International Conference on
  Intelligent User Interfaces}, page 410–422, Sydney NSW Australia. ACM.

\bibitem[\protect\astroncite{Schuff et~al.}{2011}]{Schuff_Corral_Turetken_2011}
Schuff, D., Corral, K., and Turetken, O. (2011).
\newblock Comparing the understandability of alternative data warehouse
  schemas: An empirical study.
\newblock {\em Decision Support Systems}, 52(1):9–20.

\bibitem[\protect\astroncite{Slack
  et~al.}{2019}]{Slack_Friedler_Scheidegger_Roy_2019}
Slack, D., Friedler, S.~A., Scheidegger, C., and Roy, C.~D. (2019).
\newblock Assessing the local interpretability of machine learning models.
\newblock (arXiv:1902.03501).
\newblock arXiv:1902.03501.

\bibitem[\protect\astroncite{Spatola}{2024}]{Spatola_2024}
Spatola, N. (2024).
\newblock The efficiency-accountability tradeoff in ai integration: Effects on
  human performance and over-reliance.
\newblock {\em Computers in Human Behavior: Artificial Humans}, page 100099.

\bibitem[\protect\astroncite{Tai
  et~al.}{2024}]{Tai_Bentley_Xia_Sitt_Fankhauser_Chicas-Mosier_Monteith_2024}
Tai, R.~H., Bentley, L.~R., Xia, X., Sitt, J.~M., Fankhauser, S.~C.,
  Chicas-Mosier, A.~M., and Monteith, B.~G. (2024).
\newblock An examination of the use of large language models to aid analysis of
  textual data.
\newblock {\em International Journal of Qualitative Methods},
  23:16094069241231168.

\bibitem[\protect\astroncite{Vasconcelos et~al.}{2023}]{Vasconcelos_2023}
Vasconcelos, H., Jörke, M., Grunde-McLaughlin, M., Gerstenberg, T., Bernstein,
  M.~S., and Krishna, R. (2023).
\newblock Explanations can reduce overreliance on ai systems during
  decision-making.
\newblock {\em Proceedings of the ACM on Human-Computer Interaction},
  7(CSCW1):1–38.

\bibitem[\protect\astroncite{Vered et~al.}{2023}]{Vered_Livni2023}
Vered, M., Livni, T., Howe, P. D.~L., Miller, T., and Sonenberg, L. (2023).
\newblock The effects of explanations on automation bias.
\newblock {\em Artificial Intelligence}, 322:103952.

\bibitem[\protect\astroncite{Wang et~al.}{2019}]{Wang_Yang_Abdul_Lim_2019}
Wang, D., Yang, Q., Abdul, A., and Lim, B.~Y. (2019).
\newblock Designing theory-driven user-centric explainable ai.
\newblock In {\em Proceedings of the 2019 CHI Conference on Human Factors in
  Computing Systems}, page 1–15, Glasgow Scotland Uk. ACM.

\bibitem[\protect\astroncite{Wang and Yin}{2021}]{Wang_Yin_2021}
Wang, X. and Yin, M. (2021).
\newblock Are explanations helpful? a comparative study of the effects of
  explanations in ai-assisted decision-making.
\newblock In {\em 26th International Conference on Intelligent User
  Interfaces}, IUI ’21, page 318–328, New York, NY, USA. Association for
  Computing Machinery.

\bibitem[\protect\astroncite{Wischnewski
  et~al.}{2023}]{Wischnewski_Krämer_Müller_2023}
Wischnewski, M., Krämer, N., and Müller, E. (2023).
\newblock Measuring and understanding trust calibrations for automated systems:
  A survey of the state-of-the-art and future directions.
\newblock In {\em Proceedings of the 2023 CHI Conference on Human Factors in
  Computing Systems}, CHI ’23, page 1–16, New York, NY, USA. Association
  for Computing Machinery.

\bibitem[\protect\astroncite{Yates and Potworowski}{2012}]{yates2012evidence}
Yates, J.~F. and Potworowski, G.~A. (2012).
\newblock Evidence-based decision management.

\bibitem[\protect\astroncite{You et~al.}{2022}]{You_Yang_Li_2022}
You, S., Yang, C.~L., and Li, X. (2022).
\newblock Algorithmic versus human advice: Does presenting prediction
  performance matter for algorithm appreciation?
\newblock {\em Journal of Management Information Systems}, 39(2):336–365.

\bibitem[\protect\astroncite{Zhang et~al.}{2020}]{Zhang_Liao_Bellamy_2020}
Zhang, Y., Liao, Q.~V., and Bellamy, R. K.~E. (2020).
\newblock Effect of confidence and explanation on accuracy and trust
  calibration in ai-assisted decision making.
\newblock In {\em Proceedings of the 2020 Conference on Fairness,
  Accountability, and Transparency}, FAT* ’20, page 295–305, New York, NY,
  USA. Association for Computing Machinery.

\end{thebibliography}

\appendix
\pagenumbering{roman} 
\section{Instructions}
\label{appendix:instructions}

Dear Participant,

Thank you for your interest in our study. This page provides you with detailed instructions to guide you through the study. Please read them carefully before you begin.

\textbf{Study Overview}\\
This study focuses on income estimation, where you assess whether a person's net income is above the median income. The median income is the point at which half of the employed population earns more and the other half earns less. In this study, "above the median" means that the person's income belongs to the richer half of the population.

This involves individuals from Germany. The median income is €1615 net per month. This includes all employed persons over 18 years old who are not receiving a pension.

You will participate in 4 rounds. In each round, you will receive information about a real person. Your task is to estimate the probability, using percentages, that this person's income is above the median.

At the end of the study, your estimates will be compared with the actual data to determine whether the person's income is indeed higher than the median. Based on this comparison, you will receive a bonus payment. The bonus is calculated using the so-called Brier Score. For example, if you always say the probability is 50\%, the Brier Score is 0.25, and in this case, you will not receive a bonus. The better your probability estimates, the smaller the Brier Score and the higher your bonus payment. With a Brier Score of 0, you have estimated perfectly and will receive a bonus of £6. Regardless of your performance, you will receive a fixed compensation of £3 for participating in the study.

[ if treatment is not \co ]

\textbf{Artificial Intelligence (AI) Support}\\
You will receive support from an Artificial Intelligence (AI) for your income estimates. The AI was trained using data from over 1,500 individuals to estimate as accurately as possible whether their income is above the median. The AI is not perfect; it is correct 77\% of the time.

[ if treatment is \ro ] The AI will provide you with recommendations on the probability that each person's income is above the median. For example, the AI might say that it believes the probability is 65\%. [ endif ]

[ if treatment is \eo or treatment is \eva ] The AI will provide you with arguments for (pro) and against (contra) each person's income potential to assist you in your estimation.[ if treatment is \eva ] You can open the arguments with the respective buttons. [ endif ] [ endif ]

[ if treatment is \rae ] The AI will provide you with recommendations on the probability that each person's income is above the median. For example, the AI might say that it believes the probability is 65\%. Additionally, it will provide you with arguments for (pro) and against (contra) the income potential to assist you in your estimation. [ endif ]

[ if treatment is not \ro ] The AI bases its arguments on its learned knowledge and the characteristics of the evaluated individuals. For each characteristic, the AI indicates whether it is more likely to lead to an income above the median (positive arguments) or more likely to lead to an income below the median (negative arguments). Each characteristic of the individuals is rated with a number. The more positive the number, the more the AI views the characteristic as conducive to an income above the median. Conversely, the more negative the number, the more the AI views the characteristic as conducive to an income below the median. These numbers are displayed separately in bar charts. Additionally, below each chart, there is a text that briefly explains the arguments.
[ endif ]
[ endif ]

\textbf{Survey}
After completing all task rounds, you will be asked to fill out a survey.

\section{Screenshots}
\label{appendix:screenshots}

\begin{figure}[h]
    \centering
    \includegraphics[width=1\linewidth]{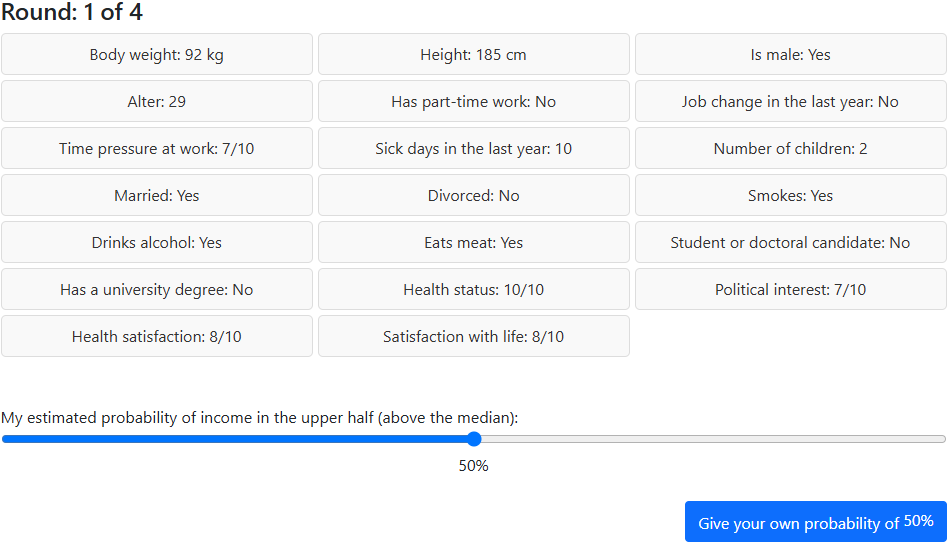}
    \caption{Translated interface in \co: The features with their values are listed at the top, followed by the input field for the participant below.}
    \label{fig:screenshot_co}
\end{figure}

\begin{figure}[h]
    \centering
    \includegraphics[width=1\linewidth]{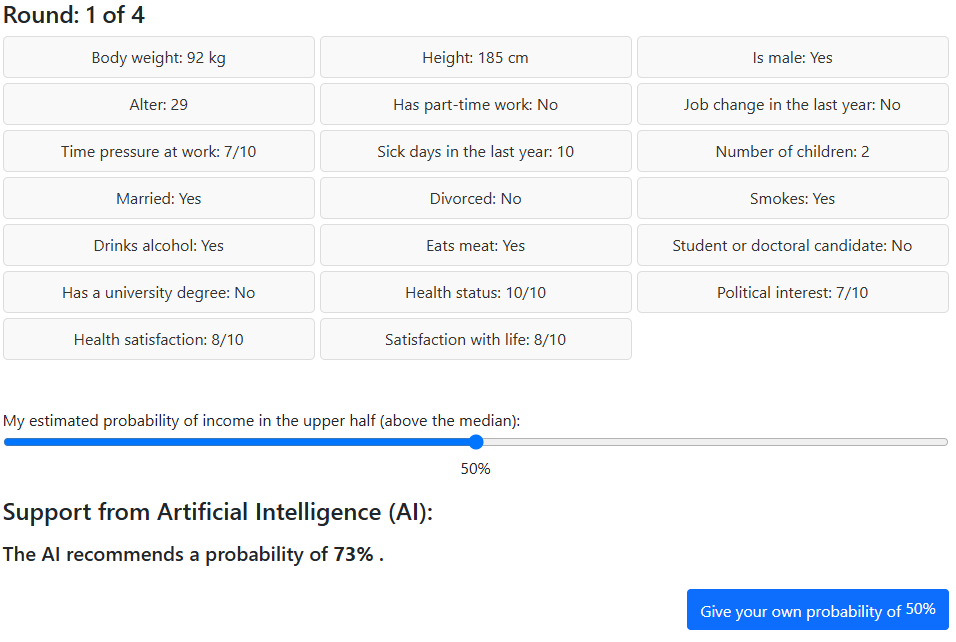}
    \caption{Translated interface in \ro: The features with their values are listed at the top, followed by the input field for the participant, and below that, the AI recommendation.}
    \label{fig:screenshot_ro}
\end{figure}

\begin{figure}[h]
    \centering
    \includegraphics[width=1\linewidth]{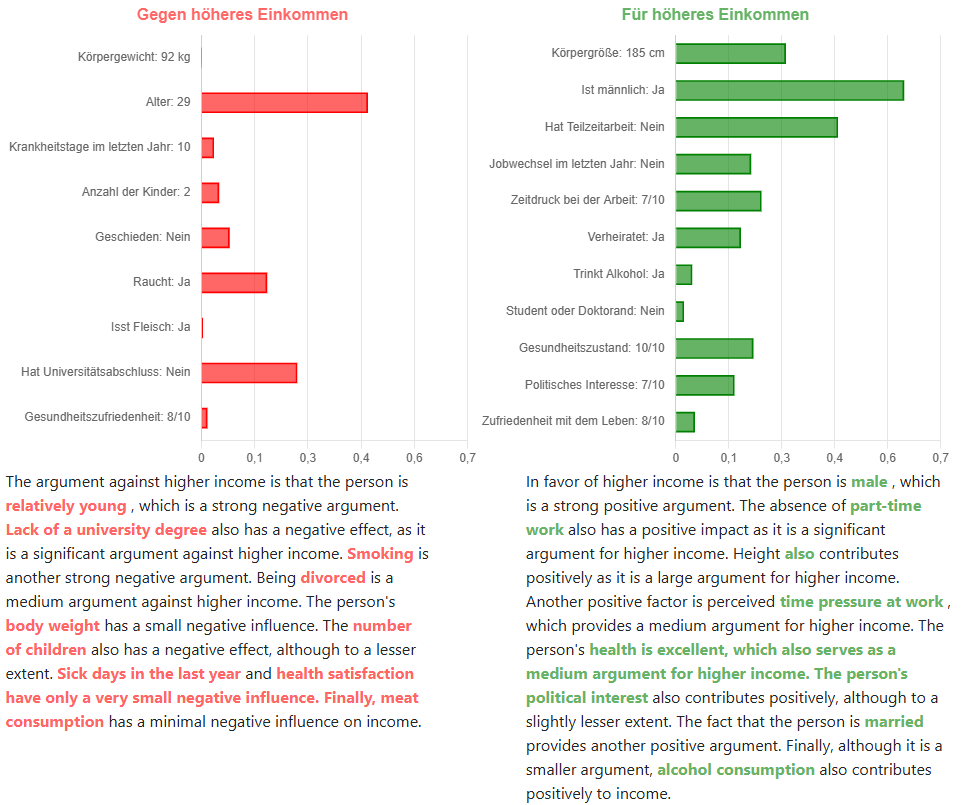}
    \caption{Translated interface in \eo: Due to space constraints in the screenshot, the features and input field were not included, only the presentation of the pro and con evidence.}
    \label{fig:screenshot_eo}
\end{figure}

\begin{figure}[h]
    \centering
    \includegraphics[width=1\linewidth]{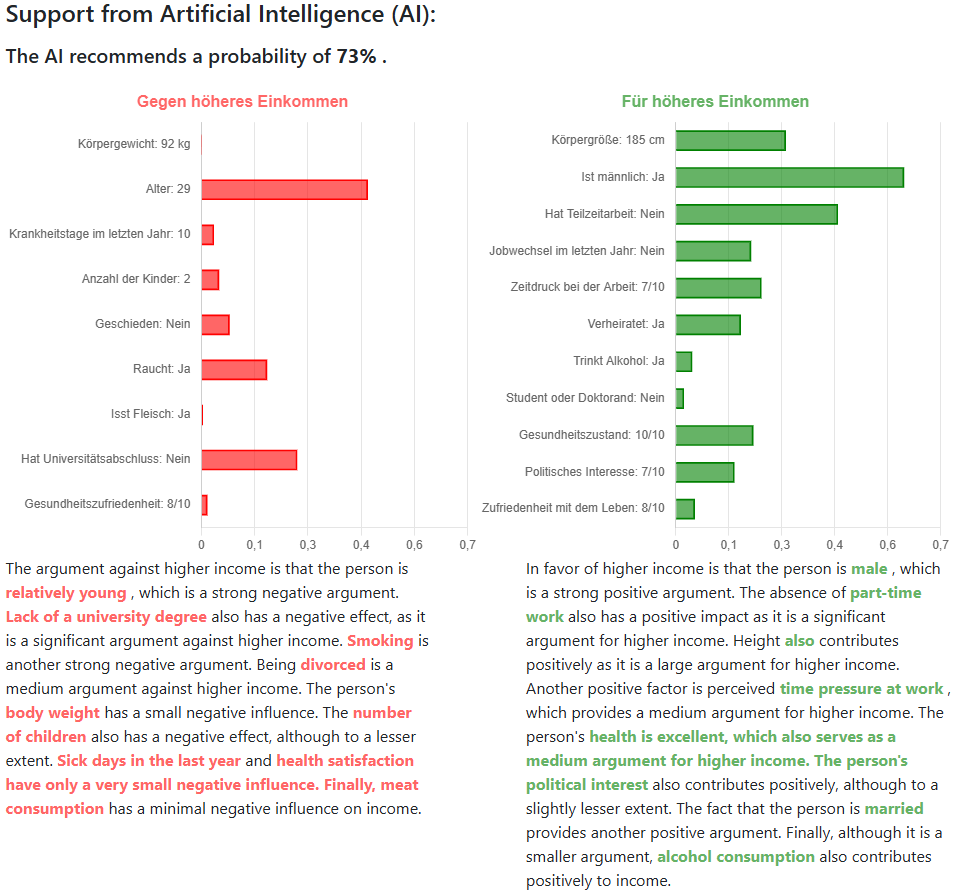}
    \caption{Translated interface in \rae: Due to space constraints in the screenshot, the features and input field were not included, only the recommendation and the presentation of the pro and con evidence.}
    \label{fig:screenshot_rae}
\end{figure}

\begin{figure}[h]
    \centering
    \includegraphics[width=1\linewidth]{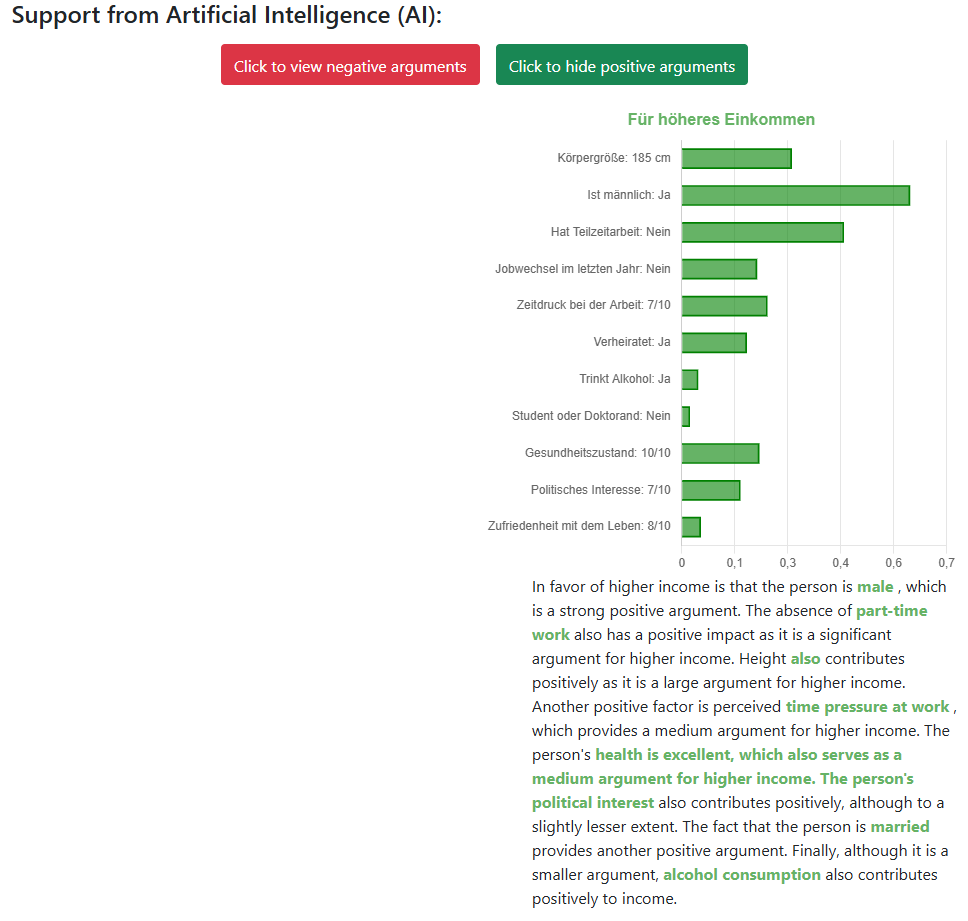}
    \caption{Translated interface in \eva: Due to space constraints in the screenshot, the features and input field were not included, only the button for the con evidence and the already displayed pro evidence.}
    \label{fig:screenshot_eva}
\end{figure}

\end{document}